\documentclass[final,english,twocolumn,amssymb, nobibnotes, aps, prx, longbibliography]{revtex4-1}
\usepackage{graphicx}
\usepackage{natbib}
\usepackage{amsmath}
\usepackage{amssymb}
\usepackage{bbold}
\usepackage[caption=false]{subfig}
\usepackage{float}
\usepackage{appendix}
\usepackage[dvipsnames]{xcolor}
\usepackage[section]{placeins}
\definecolor{darkblue}{rgb}{0,0,.65}
\definecolor{darkgreen}{rgb}{0.28,0.41,0.19}

\usepackage{wrapfig}

\usepackage[%
pdfauthor={David J. Luitz},%
pdfstartview=FitH,%
breaklinks=true,%
bookmarks=true,%
colorlinks=true,%
anchorcolor=black,%
citecolor=blue,
filecolor=black,%
menucolor=black,%
urlcolor=darkblue,%
linkcolor=blue,%
]{hyperref}
\usepackage[all]{hypcap} %
\newcommand{\bra}[1]{\langle\,#1\,|}
\newcommand{\ket}[1]{|#1\rangle}

\newcommand{\braket}[2]{\langle\,#1\, | \, #2\,\rangle}

\newcommand{\E}{\mathrm{e}}

\newcommand{\D}{\mathrm{d}}

\newcommand{\up}{\uparrow}
\newcommand{\dw}{\downarrow}

\graphicspath{{images/}}
\begin{document}

	\title{Three-dimensional isometric tensor networks}
	
	\author{Maurits S. J. Tepaske}
	\email{mtepaske@pks.mpg.de}
	\author{David J. Luitz}
	\email{dluitz@pks.mpg.de}
	\affiliation{Max Planck Institute for the Physics of Complex Systems, Noethnitzer Str. 38, 01167 Dresden, Germany}
	\date{\today}
	
	\begin{abstract}
		Tensor network states are expected to be good representations of a large class of interesting quantum many-body wave functions. In higher dimensions, their utility is however severely limited by the difficulty of contracting the tensor network, an operation needed to calculate quantum expectation values. Here we introduce a method for the time-evolution of three-dimensional isometric tensor networks which respects the isometric structure and therefore renders contraction simple through a special canonical form. Our method involves a tetrahedral site-splitting which allows to move the orthogonality center of an embedded tree tensor network in a simple cubic lattice to any position.
		
		Using imaginary time-evolution to find an isometric tensor network representation of the ground state of the 3D transverse field Ising model across the entire phase diagram, we perform a systematic benchmark study of this method in comparison with exact Lanczos and quantum Monte Carlo results. We show that the obtained energy matches the exact groundstate result accurately deep in the ferromagnetic and polarized phases, while the regime close to the critical point requires larger bond dimensions. This behavior is in close analogy with the two-dimensional case, which we also discuss for comparison.
	\end{abstract}
	\maketitle
	
	\section{Introduction}
	
	The Hilbert space dimension of quantum many-body systems grows exponentially with the number of constituents, making the direct handling of many-body wavefunctions impractical for large systems. Tensor networks are an attempt to tame the many-body wavefunction, by expressing it in terms of local tensors, which are contracted according to the network structure. This reduces the complexity from an exponential to a polynomial number of variables. While in principle any wavefunction can be expressed as a tensor network, some particularly entangled states require exponentially large tensors. Fortunately, the manifold of wavefunctions expressible with small tensor networks includes wavefunctions with area law entanglement, which are expected to be relevant for the description of ground-states of many local quantum many-body systems \cite{hastings_area_2007,hastings_locality_2004,eisert_colloquium:_2010}.
	
\begin{figure}[t]
		\begin{center}
			\includegraphics[width=0.6\linewidth]{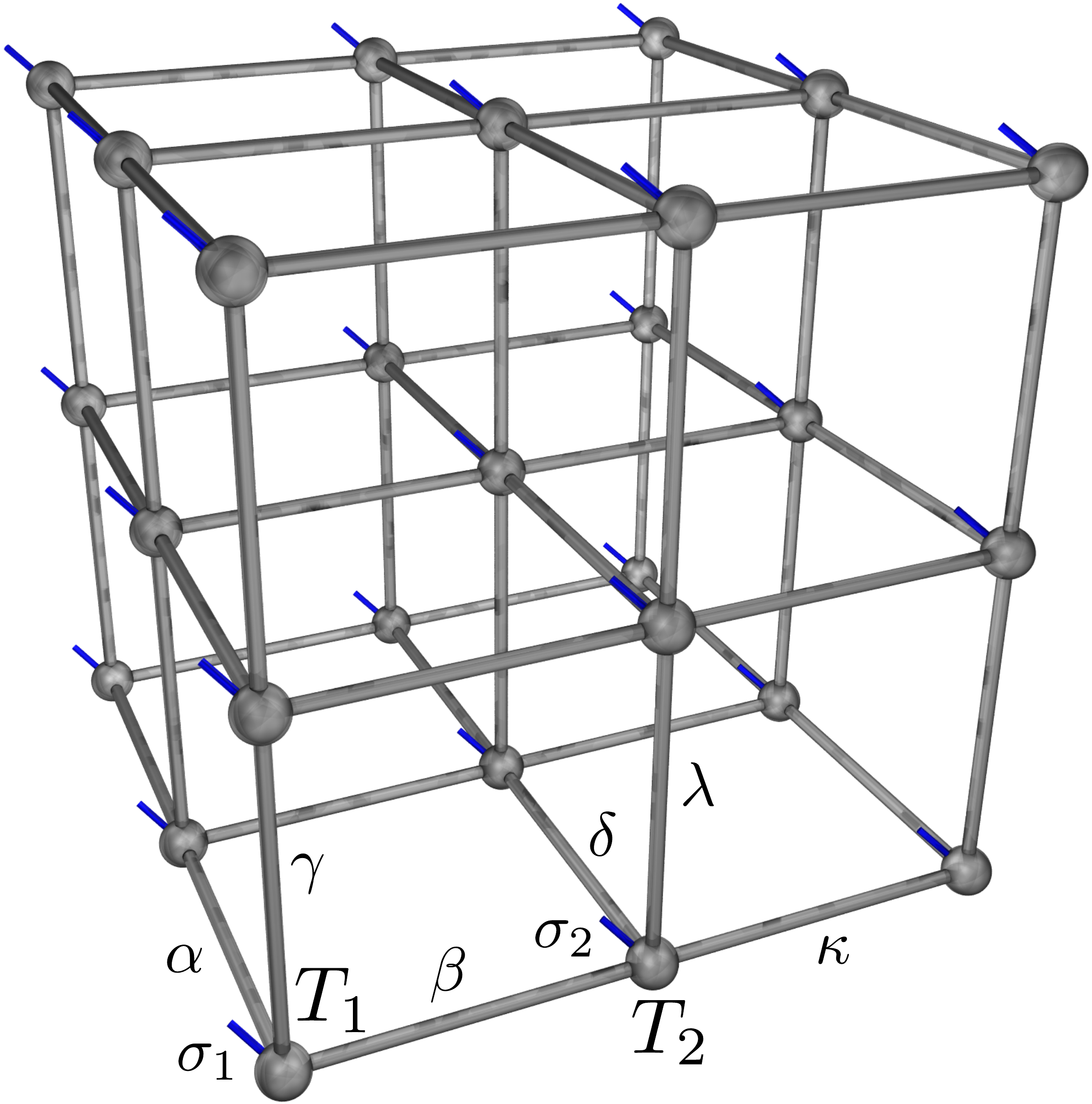}
		\end{center}
		\caption{A generic 3D PEPS ansatz for the cubic lattice, where the tensors $T^{\sigma_i}_{i\alpha\beta\gamma\delta\kappa\lambda}$ are represented by spheres. The blue legs denote the physical degrees of freedom $\sigma_i$ and the gray legs denote the virtual degrees of freedom. The connections depict contractions between the virtual legs of neighboring tensors.}\label{fig:1}
	\end{figure}

	Tensor network states are particularly successful in one dimension (1D), where they are known as "matrix-product states" (MPS) \cite{SCHOLLWOCK201196}, which have become state-of-the-art machinery for the classical simulation of 1D many-body systems. This popularity rests primarily on the existence of powerful algorithms to variationally optimize the energy of the state (e.g. the density matrix renormalization group \cite{white_density_1992}) and on the ability to compute matrix elements of local operators $\bra{\phi} \hat O \ket{\psi}$ both exactly and efficiently. In particular the second property does not generalize to the higher-dimensional variants of MPS known as "projected-entangled pair states" (PEPS) \cite{verstraete2004renormalization}. It turns out that the exact calculation of a local correlator in an arbitrary PEPS state $\bra{\text{PEPS}}\hat {O}\ket{\text{PEPS}}$ -- requiring the contraction of a higher-dimensional network -- is generally inefficient for generic finite PEPS with open boundary conditions (OBC) already in two dimensions \cite{verstraete2004renormalization, PhysRevLett.98.140506, Lubasch_2014, doi:10.1080/14789940801912366, PhysRevB.90.064425}. While PEPS are readily formulated in three dimensions (cf. Fig. \ref{fig:1}), currently no efficient contraction method is known. So even though PEPS are efficient representations of area-law entangled quantum many-body wavefunctions, it is often difficult to extract useful information from them.
	
	The central problem for the generalization of powerful 1D methods to higher dimensions is caused by the fact that cutting a bond in a higher-dimensional PEPS does not separate the network into two disconnected pieces, in contrast to 1D MPS. In MPS methods, the separation of the network into unique "left" and "right" parts by cutting any bond is exploited by using an orthonormal basis to represent the left/right states, and one can then decimate the basis to the dominant components by truncating to the largest singular values in an optimal way \cite{vidal_efficient_2003}. This property is the foundation of MPS evolution algorithms \cite{hubig_review}. The absence of such separability in higher-dimensional PEPS diminishes the effectiveness of purely local evolution algorithms, where in the case of a nearest-neighbor interacting system the tensor network is optimized by iterating over the bonds and applying a two-body gate to each bond followed by a truncation of this bond according to the standard time-evolving block-decimation (TEBD) \cite{SCHOLLWOCK201196, vidal_efficient_2003}. Instead, optimal truncation and hence optimal evolution requires each gate to be accompanied by a contraction of the full network (dubbed "full update"), which is inefficient as it generally scales exponentially with network size when performed exactly \cite{Lubasch_2014}.
	
	One way around the inefficiency of full contraction is to instead perform the contraction approximately \cite{verstraete2004renormalization, Lubasch_2014, doi:10.1080/14789940801912366, PhysRevB.90.064425}, sacrificing precision for speed.
	Recently there appeared multiple works \cite{zaletel2019isometric, MZtalk, PhysRevB.100.054404, hyatt2019dmrg} which suggest an attractive alternative: to construct finite PEPS in an \textit{explicit canonical form} in which it can be contracted both exactly and efficiently in a way that local truncation again becomes optimal just like for MPS, thereby circumventing the mentioned problems that occur when dealing with generic PEPS. This does induce a loss of generality, restricting its subspace in Hilbert space to a subspace of generic PEPS, thereby reducing the expressivity of the network \cite{soejima2019isometric}. While the effect of this restriction is not yet clear, it becomes irrelevant in the limit of large bond dimensions and therefore seems at least in principle controllable.
	
	In \cite{zaletel2019isometric} a class of finite 2D PEPS called "isometric tensor network states" (isoTNS) was introduced for which $\braket{\text{PEPS}}{\text{PEPS}}$ reduces to a canonical MPS norm, and which can be time-evolved using an efficient local evolution algorithm called $\text{TEBD}^2$. Here we will generalize the isoTNS ansatz to 3D and develop an extension of $\text{TEBD}^2$ which we call $\text{TEBD}^3$. This upgrade to a higher spatial dimension is an important step in developing efficient techniques to simulate generic 3D quantum many-body systems, especially for cases which are not accessible to quantum Monte Carlo methods due to a sign-problem. This importance is stressed by the limited number of existing finite 3D PEPS algorithms \cite{TENG2017117, PhysRevB.87.085130} and generic simulation methods for 3D quantum many-body systems in general \cite{PhysRevLett.97.107206, PhysRevB.81.214426, SciPostPhys.4.2.013}.

	\section{Method}
	
	A generic finite 3D PEPS ansatz for a 3D many-body spin-1/2 system can be written in the local basis $\sigma_i=\pm1$ as
	\begin{equation}
	\ket{\text{PEPS}} = \sum_{\sigma_1...\sigma_N}\mathcal{C}\left(T^{\sigma_1}_1...T^{\sigma_N}_N\right)\ket{\sigma_1...\sigma_N},
	\end{equation}
	where $T^{\sigma_i}_i$ represents the set of tensors which contain the complex-valued variational parameters and which are spatially arranged like the spins $\sigma_i$. Here $\mathcal{C}$ indicates that all tensors are contracted, giving complex scalar coefficients, which is usually done by choosing the amount of virtual degrees of freedom per $T^{\sigma_i}_i$ equal to the lattice connectivity and then contracting nearest-neighbors. 
	In Fig. \ref{fig:1} this is illustrated for a cubic lattice with open boundary conditions, where the pairs of virtual degrees of freedom are represented by the gray bonds and the physical (spin) degrees of freedom are represented by the blue free legs, i.e. in a particular basis we get the tensors $T^{\sigma_i}_{i\alpha\beta\gamma\delta\kappa\lambda}$.
	
	In order to calculate $\braket{\text{PEPS}}{\text{PEPS}}$ we would contract this tensor network with the physical legs of its conjugate, which we would then have to contract down to a scalar.

	\subsection{General properties of isoTNS}

	The goal of this work is to design a type of 3D tensor network that allows for the full network contraction to be done exactly and efficiently. 
	To this end we impose an additional structure on the PEPS shown in Fig. \ref{fig:1}, such that $\braket{\text{PEPS}}{\text{PEPS}}=\braket{\text{MPS}}{\text{MPS}}$ becomes manifest. In particular, we choose the majority of the $T_i$ to be isometric, meaning that these $T_i$ reduce to identities when contracted with their conjugate over a subset of the legs, e.g.
	\begin{equation}
	\label{eq:2}
	\sum_{i\alpha\beta\gamma\delta\kappa}T_{\sigma_i}^{\dagger\alpha\beta\gamma\delta\kappa\lambda}T_{\sigma_i}^{\alpha\beta\gamma\delta\kappa\eta}=\mathbb{1}^{\lambda\eta},
	\end{equation}
	and
	\begin{equation}
	\label{eq:3}
	\sum_{i\alpha\beta\gamma\delta}T_{\sigma_i}^{\dagger\alpha\beta\gamma\delta\kappa\lambda}T_{\sigma_i}^{\alpha\beta\gamma\delta\nu\eta}=\mathbb{1}^{\kappa\nu}\mathbb{1}^{\lambda\eta}.
	\end{equation}
	If $T$ is unitary instead of isometric, these constraints also hold under $T\rightarrow T^{\dagger}$, which in the case of an isometry instead gives a projector.
	
	\begin{figure}[H]
		\begin{center}
			\includegraphics[width=\columnwidth]{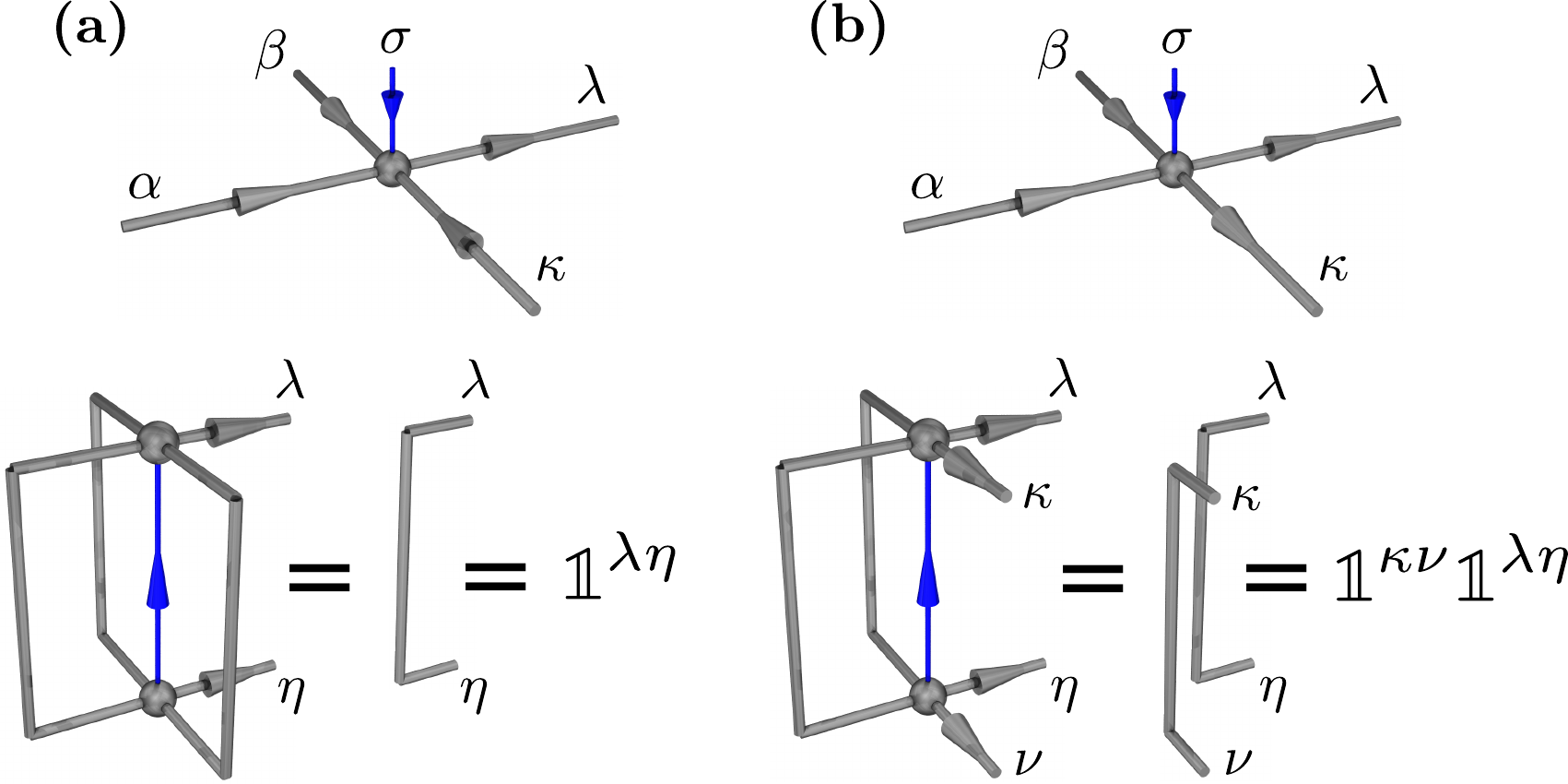}
		\end{center}
		\caption{The isometry constraints are encoded by decorating the tensor legs with arrows (top row). Here panel (a) corresponds to Eq. \ref{eq:2} and panel (b) to Eq. \ref{eq:3}, where in the bottom figures the tensor shown in the top panel is contracted with its conjugate to yield identities.}\label{fig:2}
	\end{figure}
	
	\begin{figure*}[t!]
		\begin{center}
			\includegraphics[width=0.9\textwidth]{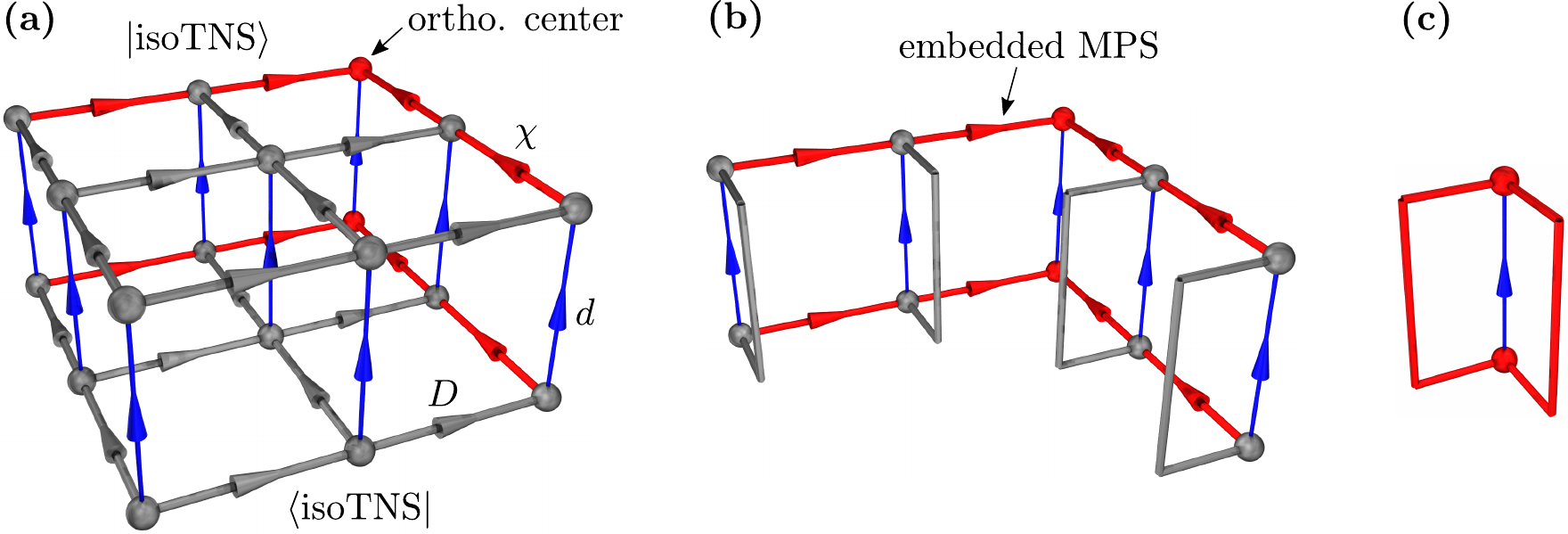}
		\end{center}
		\caption{Multiple stages of the reduction of $\braket{\text{isoTNS}}{\text{isoTNS}}$ to a contraction of just two tensors, illustrated for a 2D isoTNS on a $3\times3$ square lattice. Panel (a) shows the amplitude before employing the isometry constraints, where the gray bonds have dimension $D$, the red bonds have dimension $\chi$, and the blue bonds have dimension $d$. Panel (b) shows the canonical MPS that is obtained after utilizing a large part of the isometry structure. Panel (c) shows the final pair of contracted tensors that remains after fully utilizing the isometry structure.}\label{fig:3}
	\end{figure*}

	In the language of tensor network diagrammatics we can represent these constraints by decorating the legs with arrows \cite{SCHOLLWOCK201196, zaletel2019isometric}, where incoming arrows represent contracted indices in the isometry constraint and where outgoing arrows represent free indices. In Fig. \ref{fig:2} we illustrate this notation for tensors with four virtual legs. Here the diagrams in the upper panels encode the isometry constraints depicted in the lower panels. Specifically, Fig. \ref{fig:2}a corresponds to Eq. \ref{eq:2} and Fig. \ref{fig:2}b to Eq. \ref{eq:3}. Note that for convenience we choose a single arrow direction after contracting the legs. In the same spirit we will omit the arrows on physical legs from here on, since we will always choose these to be incoming.
	
	Using these isometric tensors we can construct tensor networks that identically reduce to a single pair of tensors (the "orthogonality center") upon contraction with its conjugate, which are called isoTNS \cite{zaletel2019isometric}. In Fig. \ref{fig:3} we show a few snapshots of this reduction for a 2D isoTNS on a $3\times3$ square lattice. Before employing the isometry constraints we have the network depicted in Fig. \ref{fig:3}a. Here we allow the red and gray legs to have distinct bond dimensions $\chi$ and $D$, which is motivated by the observation that for the chosen pattern of arrows we can first reduce the network to a canonicalized MPS amplitude, as shown in Fig. \ref{fig:3}b. Note that the gray legs which stem from the reduction effectively enlarge the local Hilbert space of the MPS.
	
	The distinction between red MPS bonds and gray non-MPS bonds is an important aspect of the isoTNS ansatz, as it will allow us to increase the accuracy of correlators while increasing the bond dimension on only a small part of the tensor network. In the Sec. \ref{results} we will see that in some situations the accuracy is indeed greatly increased when we increase $\chi$ while keeping $D$ constant. This is especially favorable because we will see in Sec. \ref{compcost} that the computational cost of the time-evolution algorithm scales significantly more favorably in $\chi$ than in $D$.
	
	Due to the remaining isometry structure we can further reduce the MPS amplitude down to a single site, as shown in Fig. \ref{fig:3}c. This final site, which is called the orthogonality center and which is colored red in Fig. \ref{fig:3}, therefore fully encodes the norm, just like the orthogonality center of MPS in 1D \cite{SCHOLLWOCK201196}. It also encodes the one-body correlators of that site, but if we want to calculate its two-body correlator with another site we can already no longer reduce the network to the orthogonality center. In terms of the arrow language this reduction requires that the arrows flow towards the MPS and within the MPS towards the orthogonality center, which can be characterized as having only incoming arrows.
	
\begin{figure}[h]
		\begin{center}
			\includegraphics[width=0.6\columnwidth]{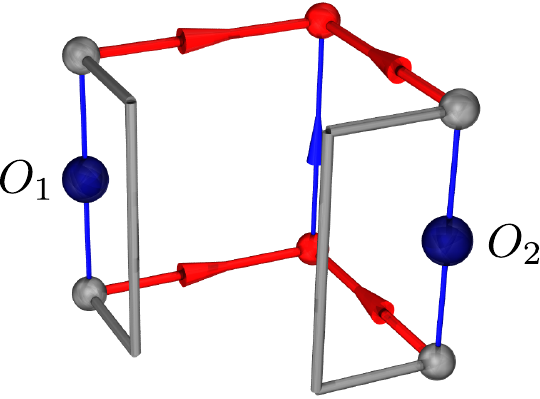}
		\end{center}
		\caption{The final product of the isometry reductions when calculating the local two-body correlator $\bra{\textrm{isoTNS}}O_1O_2\ket{\textrm{isoTNS}}$, leaving us with a MPS correlator that has to be fully computed.}\label{insertions}
	\end{figure}

	For example, say we want to calculate a local correlator between the two nearest-neighbors of the center in Fig. \ref{fig:3}b. In this case the MPS reduction that yields Fig. \ref{fig:3}c from Fig. \ref{fig:3}b would be halted at the operators, since there we can no longer utilize the isometry relations from Fig. \ref{fig:2}. Consequently we also cannot utilize the isometric tensors that lie between the operators, meaning that we are left with a MPS correlator instead of a single-site correlator, the accuracy of which is controlled by $\chi$. This is illustrated for $\bra{\textrm{isoTNS}}O_1O_2\ket{\textrm{isoTNS}}$ in Fig. \ref{insertions}.
	
	\begin{figure*}[!t]
		\begin{center}
			\includegraphics[width=0.9\textwidth]{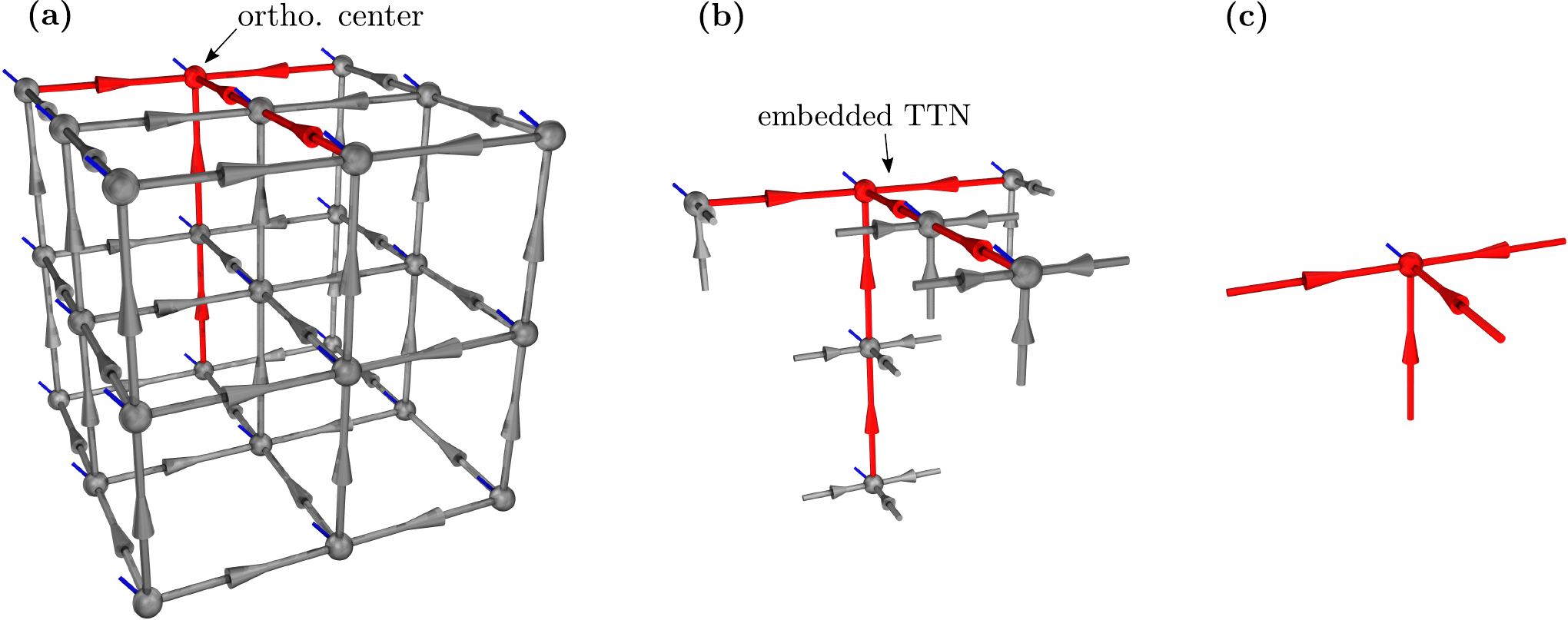}
		\end{center}
		\caption{The reduction of a particular 3D isoTNS configuration as it would occur upon contraction with its conjugate. In panel (a) we show the initial network. In panel (b) we show the canonical TTN that would remain after utilizing most of the isometry structure. In panel (c) we show how the canonical TTN further reduces to a single site, i.e. the orthogonality center.}\label{fig:4}
	\end{figure*}
	
	Clearly the calculation of a MPS correlator is more costly than the single-site correlator, but crucially it still scales polynomially in its bond dimension and size \cite{SCHOLLWOCK201196}. When we instead consider local correlators with one or more operators located outside of the MPS, the isometry reductions yield a genuine 2D network, which would have to be contracted in order to get the correlator. As a result the correlator can no longer be computed efficiently, illustrating the importance of being able to move the embedded MPS through the network.
	
	It should be noted that by calculating correlators as MPS correlators we do reduce the formal expressibility of the ansatz, as compared to generic PEPS, since it is known that MPS with finite $\chi$ can only encode exponentially decaying correlations \cite{SCHOLLWOCK201196} whereas generic PEPS can also encode algebraic correlations \cite{criticalPEPS}. This means that the isoTNS ansatz can only encode exponential correlations. Nonetheless, for finite systems the bond dimension can always be chosen large enough to encode algebraic correlations which are cut off by the system size.
	
	Before discussing the time-evolution algorithm, which we will do directly for 3D isoTNS, we consider how Fig. \ref{fig:3} generalizes to 3D. First we note that the suitable embedding even for 2D is actually a tree tensor network (TTN) with the geometry of a star \cite{foundingTTN,starTTN}, which turns into a MPS when the center occupies a corner as in Fig. \ref{fig:3}. When it instead occupies the bulk there emerge four MPS strands from the center, which becomes six for a 3D cubic lattice. As long as there are no loops we can put the TTN in canonical form. The principles that underlie the isometry reduction generalize directly to 3D, and in Fig. \ref{fig:4} we show the 3D analog of Fig. \ref{fig:3}.

	\subsection{Evolving 3D isoTNS}\label{sec:evolving3D}
	
	We will now explain how the TTN is moved through the 3D isoTNS during its trotterized time-evolution, specializing to nearest-neighbor interacting Hamiltonians. In order to take advantage of the isoTNS representation we always apply evolution gates when all sites occupy the TTN, with the orthogonality center at one of the sites. We will see that this gives rise to a threefold-nested TEBD, which we call $\text{TEBD}^3$ in analogy to $\text{TEBD}^2$ for 2D isoTNS.
	
	\begin{figure*}[!t]
		\begin{center}
			\includegraphics[width=\textwidth]{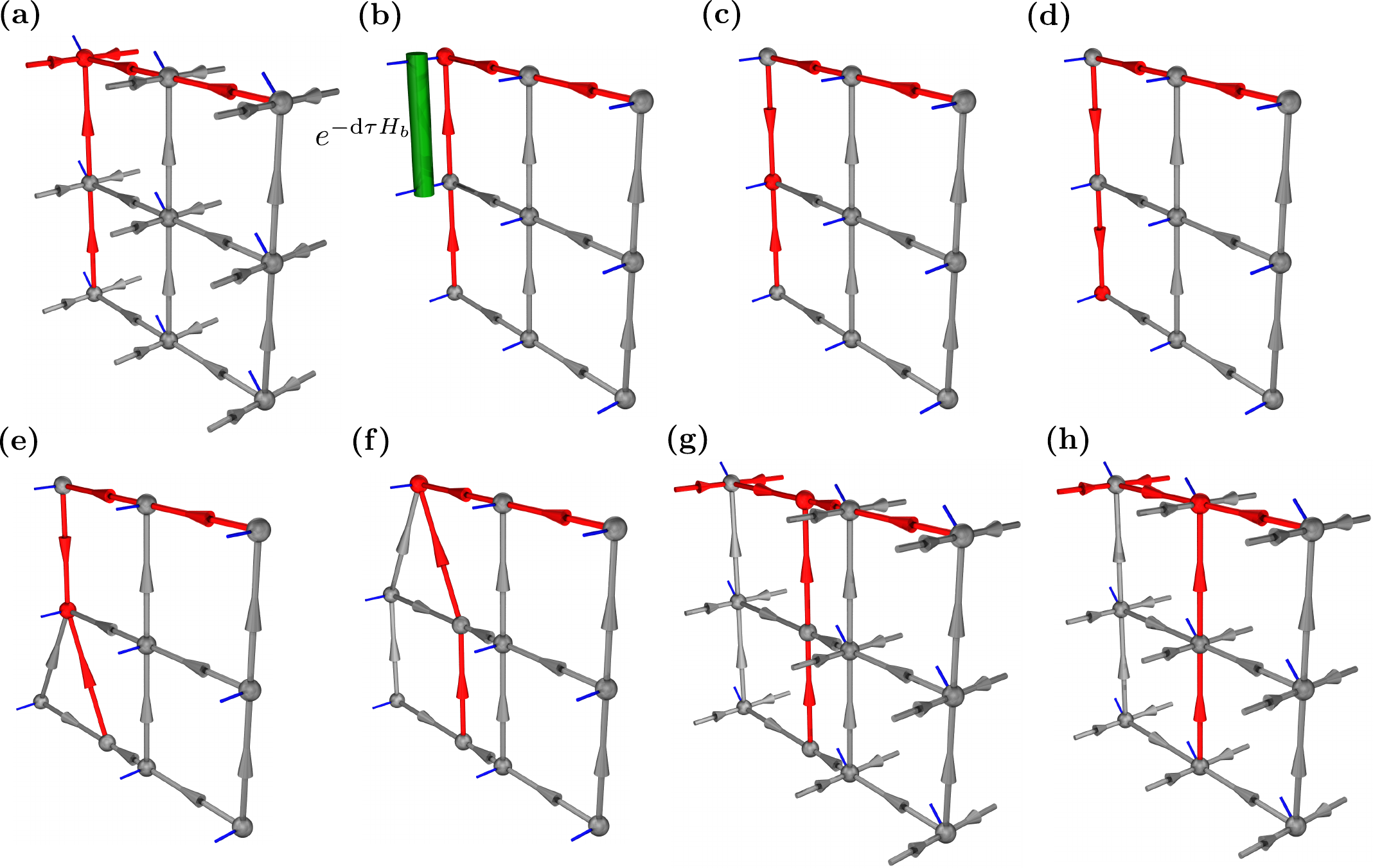}
		\end{center}
		\caption{The evolution of a slice in the 3D isoTNS, where for clarity we omit the transverse legs in most of the panels. In panel (a) we show the initial slice, which is the middle slice from Fig. \ref{fig:4}. In panel (b) we apply a two-body gate at the orthogonality center, after which in panel (c) we shift the orthogonality center downwards using a SVD. In panel (d) we repeat this after which the orthogonality center is at the bottom of the first column. In panel (e) we perform the triangle-splitting from Fig. \ref{fig:7} on the orthogonality center, which we repeat in panel (f). Now that we have reached the top of the column we perform a truncated SVD to finalize the column-splitting, after which we have transferred the TTN strand to a temporary column that has only virtual legs, as shown in panel (g) where we restored the transverse legs to emphasize that the virtual column has no transverse legs. By absorbing the virtual column into the neighboring isometry column, yielding the configuration in panel (h), we have finally moved the TTN strand to the middle column. The increased vertical bond dimension $D\chi$ is truncated back to $\chi$ before evolving the new column. Repeating the column-splitting and -evolution once more we will have evolved all columns in the slice.}\label{fig:6}
	\end{figure*}

	The evolution takes place column-wise, starting with the middle slice in Fig. \ref{fig:4}a that is shown separately in Fig \ref{fig:6}a. To this end we first trotterize the evolution operator $\exp(-\mathrm{d}\tau H)$, which is for an imaginary time-step of size $\mathrm{d}\tau$, in terms of columns $c_x,c_y,c_z$. At first order in $\mathrm{d}\tau$ we can trotterize as:
	\begin{equation}
	e^{-\mathrm{d}\tau H}\approx\prod_{c_x}e^{-\mathrm{d}\tau H_{c_x}}\prod_{c_y}e^{-\mathrm{d}\tau H_{c_y}}\prod_{c_z}e^{-\mathrm{d}\tau H_{c_z}},\label{focoltrot}
	\end{equation}
	where the error is $\mathcal{O}(\mathrm{d}\tau^2)$ and stems from the non-commutativity of columns that intersect. With only a bit more effort we can trotterize at second order:
	\begin{multline}
	e^{-\mathrm{d}\tau H}\approx\prod_{c_x}e^{-\frac{\mathrm{d}\tau}{2} H_{c_x}}\prod_{c_y}e^{-\frac{\mathrm{d}\tau}{2} H_{c_y}}\prod_{c_z}e^{-\mathrm{d}\tau H_{c_z}}\\
	\prod_{c_y}e^{-\frac{\mathrm{d}\tau}{2} H_{c_y}}\prod_{c_x}e^{-\frac{\mathrm{d}\tau}{2} H_{c_x}},\label{socoltrot}
	\end{multline}
	which has error $\mathcal{O}(\mathrm{d}\tau^3)$. Here subsequent $c_x$ terms can be combined when performing multiple time-steps. In Appendix \ref{app_trotter} we investigate the interplay between the trotterization error and the various truncation errors.
	
	Since we will be dealing exclusively with nearest-neighbor interacting systems, each column is further trotterized in terms of two-body gates. At first order this gives us
	\begin{equation}
	e^{-\mathrm{d}\tau H_{c_x}}=\prod_{b_i}e^{-\mathrm{d}\tau h_{\rm{b_i}}} + \mathcal{O}(\mathrm{d}\tau^2),\label{2btrot}
	\end{equation} 
	where $b_i$ labels the bonds in the column and $h_{b_i}\in\mathbb{C}^{4\times4}$ represents the Hamiltonian on $b_i$. At second order we get
	\begin{equation}
    e^{-\mathrm{d}\tau H_{c_x}}=\prod_{i=1}^{L}e^{-\frac{\mathrm{d}\tau}{2} h_{\rm{b_i}}}\prod_{i=L}^{1}e^{-\frac{\mathrm{d}\tau}{2} h_{\rm{b_i}}} + \mathcal{O}(\mathrm{d}\tau^3), \label{so2btrot}
    \end{equation} 	
	which combined with eq. \ref{socoltrot} yields an overall second-order trotterization.

	\subsubsection{Evolving the columns of a slice}\label{evol_cols}
	
\begin{figure}[h]
		\begin{center}
			\includegraphics[width=0.8\columnwidth]{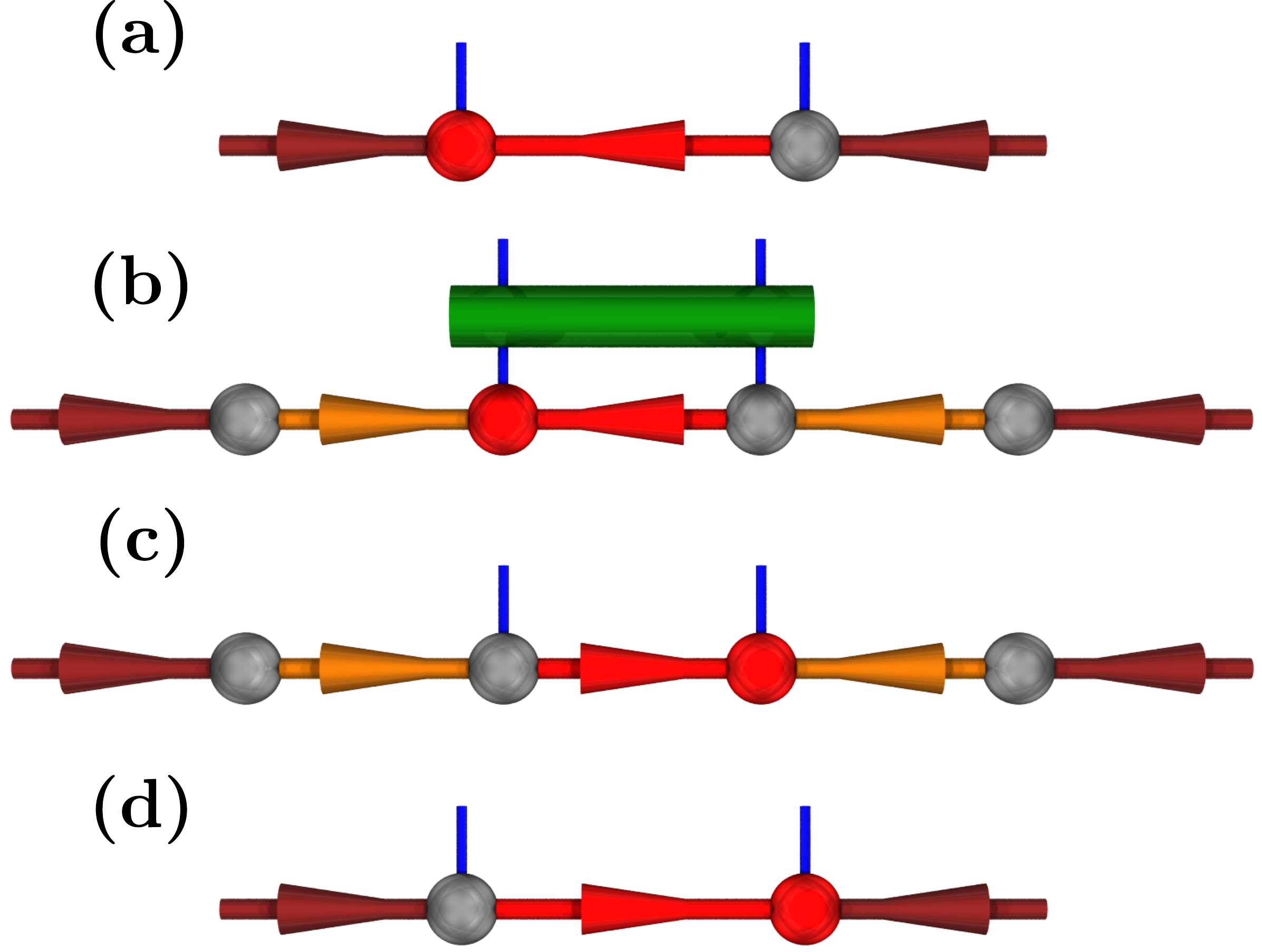}
		\end{center}
		\caption{The reduced time-evolution used in $\text{TEBD}^3$. The bond depicted in panel (a) is to be evolved, to which end we first create a reduced bond by applying a QR decomposition at each tensor. This yields a pair of transient bonds with dimension $d\chi$ that are colored orange. Because these orange bonds are much smaller than the dark-red bonds, the gate-application shown in panel (b) and the subsequent truncated SVD that yields panel (c) are much cheaper than when working in the gauge from panel (a). Having evolved the reduced bond we reabsorb the reduced tensors to get the original bond shown in panel (d), for which the orthogonality center has been shifted relative to panel (a).}\label{reduced}
	\end{figure}

	To begin the evolution we apply a two-body gate at the initial orthogonality center, as illustrated in Fig. \ref{fig:6}b. For clarity we omit the transverse legs in most of the panels, only restoring them when crucial for interpretation. Before contracting the gate with the tensors in \ref{reduced}a we apply a QR decomposition at each tensor to get a "reduced" bond \cite{PhysRevB.90.064425}, yielding the configuration in Fig. \ref{reduced}b where we now get to evolve a reduced space because the orange transient bonds are much smaller than the dark-red bonds (which consist of many virtual legs).
	
	After contracting the gate with the reduced tensors we perform a truncated singular value decomposition (SVD) $A\approx UsV^{\dagger}$, with $U$ an isometry and $sV^{\dagger}$ the new orthogonality center, to regain the reduced bond while shifting the orthogonality center, giving the configuration in Fig. \ref{reduced}c. To ensure that the new bond is not larger than the bond in Fig. \ref{reduced}a we often need to truncate the singular values $s$, which can be done optimally since we are at the orthogonality center. After reabsorbing the reduced tensors we have now shifted the orthogonality center by one site while evolving the bond, giving us the configuration in Fig. \ref{fig:6}c.
	
	We then evolve the next bond, so that we end up at the bottom as shown in Fig. \ref{fig:6}d. To evolve the next column we first need to transfer the TTN strand to this column, for which we use the column-splitting procedure introduced in \cite{zaletel2019isometric} which is sequence of triangle-splittings. This is illustrated in Fig. \ref{fig:6} (e)-(h).
	
\begin{figure}[h]
		\begin{center}
			\includegraphics[width=\columnwidth]{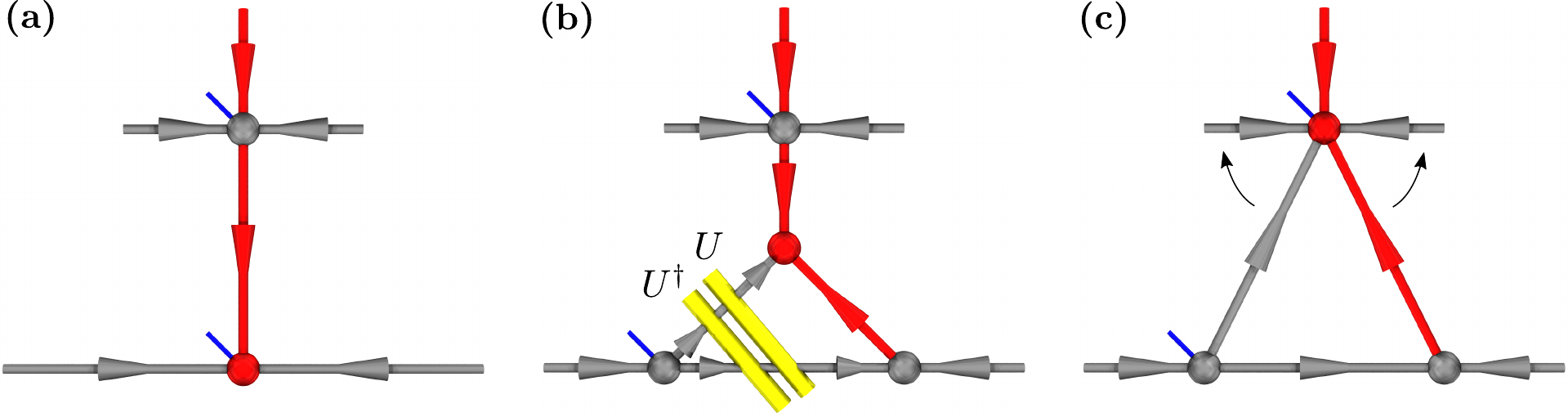}
			\caption{The triangle-splitting of the orthogonality center that is used multiple times in succession to shift a TTN strand to a neighboring column. By performing two truncated SVDs we split the orthogonality center shown in panel (a) into the triangle shown in panel (b), where the upper tensor is the new orthogonality center and where the two lower tensors are isometries. Note that the top and bottom-right tensors have only virtual legs. The amount of truncation during the SVDs is based on the usual color coding, and we have inserted a pair of disentanglers $U^{\dagger}U=\mathbb{1}$, depicted as yellow cylinders, in order to minimize the entanglement across the red bond. After the splitting we absorb the top tensor upwards, as shown in panel (c). To now perform the splitting from panel (b) on the new orthogonality center we first temporarily combine the tilted legs as indicated by the arrows, so that the orthogonality center again looks as in panel (a).}\label{fig:7}
		\end{center}
	\end{figure}

	A single triangle-splitting is shown in Fig. \ref{fig:7}. By performing two truncated SVDs on the orthogonality center in Fig. \ref{fig:7}a we get the decomposition in Fig \ref{fig:7}b. To improve the quality of the column-splitting, i.e. reduce the information-loss in obtaining Fig. \ref{fig:7}g from Fig. \ref{fig:7}d, we follow \cite{zaletel2019isometric} and reduce the bipartite entanglement between the right and upper tensors of the triangle in Fig. \ref{fig:7}b. For this we insert a pair of unitary "disentanglers" $U^{\dagger}U=\mathbb{1}$ and optimize them such that the $\alpha$-R\'enyi entropy $S_{\alpha}$ between these tensors is minimized. If we put the orthogonality center $s$ on this bond we can write
	\begin{equation}
	S_{\alpha}=\frac{1}{1-\alpha}\ln\sum_is_i^{\alpha},
	\end{equation}
	where $\alpha$ is the R\'enyi order. For $\alpha<1$ this quantity is known to provide a bound on MPS precision \cite{Verstraete_2006}. After minimizing $S_{\alpha}$ we have a triangle with minimal entanglement across its red bond, and therefore we will end up with a split-off column that has minimal vertical entanglement. The disentangler can be easily optimized with gradient descent. In the case of $\alpha=2$ there is also a cheaper optimization algorithm \cite{PhysRevB.98.235163}, but as in \cite{zaletel2019isometric} we find that $\alpha<1$ gives the best performance.
	
	After performing the triangle-splitting on the orthogonality center in Fig. \ref{fig:6}d we absorb the new center upwards as in Fig. \ref{fig:6}e and \ref{fig:7}c. To continue the column-splitting we then temporarily combine the legs as designated in Fig. \ref{fig:7}c and repeat the triangle-splitting, giving the configuration in Fig. \ref{fig:6}f, which becomes Fig. \ref{fig:6}g after a truncated SVD. This completes the column-splitting, and after absorbing the transient column into the next column we have successfully transferred the TTN strand, as shown in Fig. \ref{fig:6}h. This absorption increases the vertical bond dimension from $\chi$ to $D\chi$, which we truncate back to $\chi$ before proceeding with the evolution. After repeating the steps in Fig. \ref{fig:6} on the middle column in Fig. \ref{fig:6}h, and subsequently evolving the final column, we have moved the TTN strand from the first to the last column while evolving all columns. The final configuration is shown in Fig. \ref{fig:8}a.
	
	\subsubsection{Splitting a slice} \label{split_slice}
	
\begin{figure}[h]
		\begin{center}
			\includegraphics[width=\columnwidth]{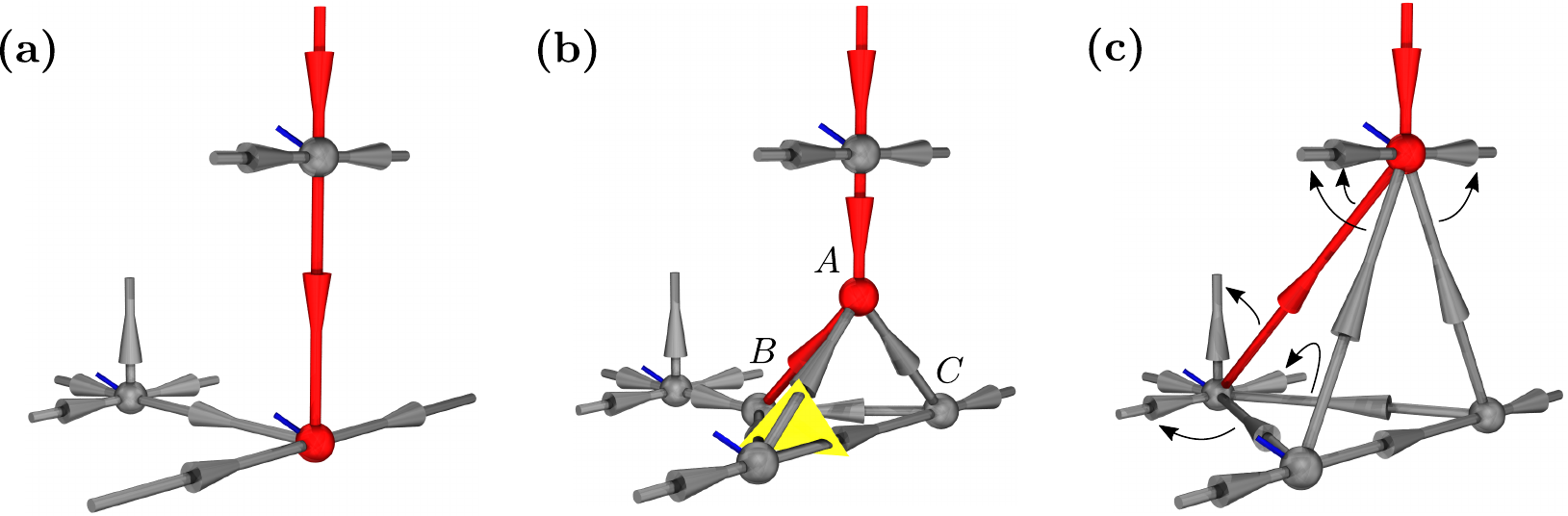}
		\end{center}
		\caption{The tetrahedron-splitting used in the slice-splitting that transfers the TTN strands to the next slice. In panel (a) we show the initial orthogonality center. In panel (b) we show the tetrahedron that results from three consecutive SVDs, where the pair of tripartite disentanglers is depicted as a yellow triangle, which reduces the tripartite entanglement $S3_{\alpha}(A|B|C)$. Note that only one of the four produced tensors has a physical leg. In panel (c) we have absorbed $A$ upwards and $B$ backwards. To repeat the tetrahedron-splitting on the new orthogonality center we temporarily combine the tilted legs as indicated by the arrows.}\label{fig:9}
	\end{figure}

\begin{figure*}[!t]
		\begin{center}
			\includegraphics[width=\textwidth]{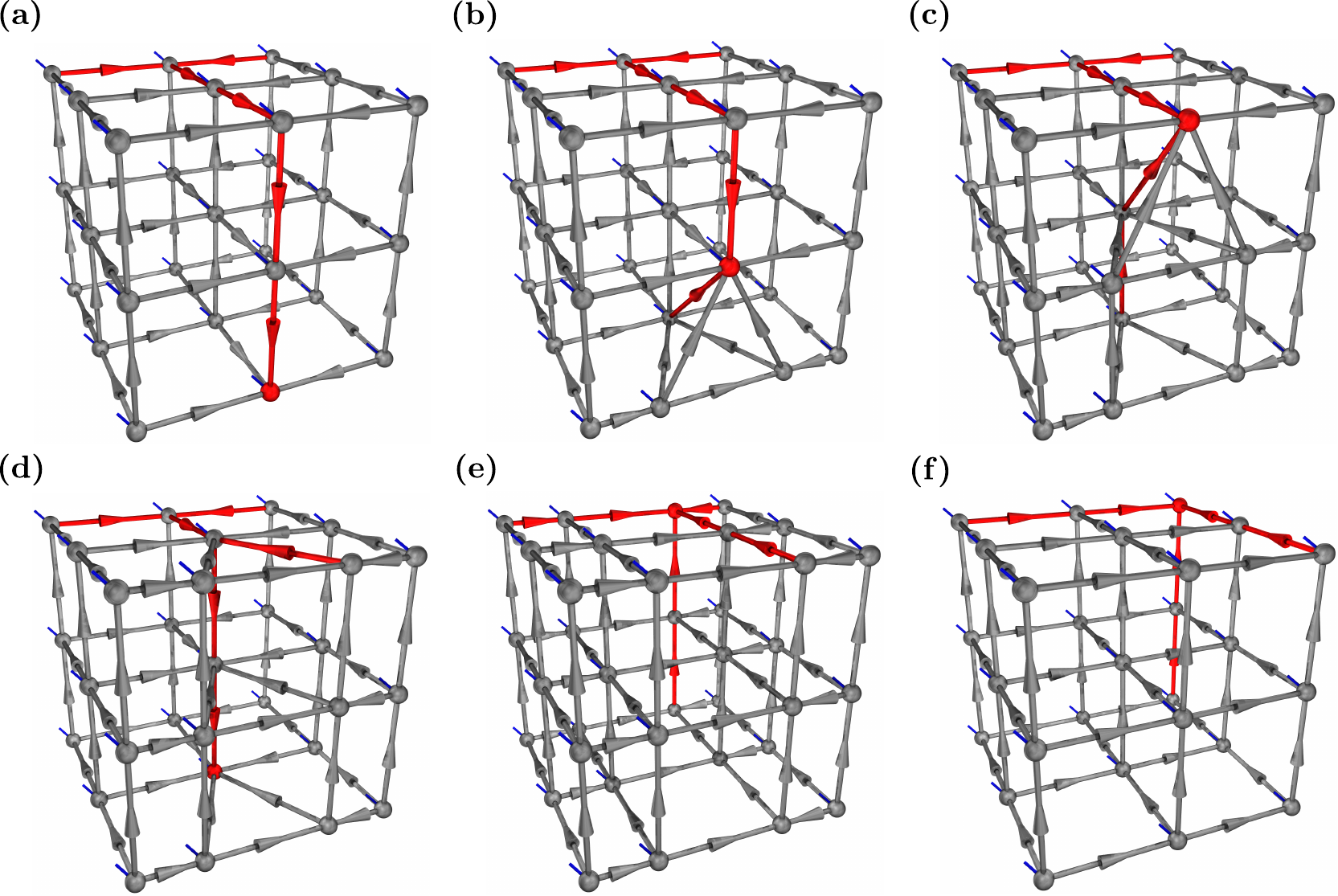}
		\end{center}
		\caption{The slice-splitting by which we transfer the two TTN strands to the next slice, after evolving the vertical bonds of a slice. In panel (a) we show the initial isoTNS, which is obtained from Fig. \ref{fig:4} by evolving the middle slice as in Fig. \ref{fig:6}. In panel (b) we show the first step of the slice-splitting, where we employ the tetrahedron-splitting from Fig. \ref{fig:9} on the orthogonality center, thereby moving it upwards by one site. After temporarily combining the tilted legs with the horizontal legs as indicated by the arrows in Fig. \ref{fig:9}c we repeat the tetrahedron-splitting to get panel (c). Finally we apply the triangle-splitting from Fig. \ref{fig:7} to the uppermost tensor in panel (d). We then move the orthogonality center to the bottom of the column, resulting in panel (d). We repeat this combination of evolving and column-splitting until we reach the final column, which is instead split via the procedure described in panels (e)-(h) of Fig. \ref{fig:6}, yielding the split-off slice in panel (e). To finalize the transfer we absorb the transient slice into the neighboring slice, yielding panel (f). The increased bond dimensions are truncated before continuing the evolution.}\label{fig:8}
	\end{figure*}

	Having evolved the columns in the first slice we now want to transfer the TTN strands to the next slice and repeat the evolution. To achieve this we use a tetrahedron-splitting on the orthogonality center, which is illustrated in Fig. \ref{fig:9}. By using a series of such splittings we can split off a transient slice containing the TTN strands, converting Fig. \ref{fig:8} into Fig. \ref{fig:8}e. Starting at the orthogonality center in Fig. \ref{fig:8}a, shown separately in Fig. \ref{fig:9}a, we perform a sequence of SVDs to get the tetrahedron in Fig. \ref{fig:9}b. The face constituted by the tensors $A,B,C$ is part of the new slice. 
	
	To improve the fidelity of the slice-splitting we now need to consider the tripartite entanglement between $A,B,C$. Hence we insert a pair of tripartite disentanglers $U^{\dagger}U=\mathbb{1}$ that minimizes a tripartite extension of the $\alpha$-R\'enyi entropy $S_{\alpha}$ (similar to the tripartite information $\mathrm{I3}$ \cite{Rangamani_2015}):
	\begin{equation}
	\mathrm{S3}_{\alpha}(A|B|C)=S_{\alpha}(A|BC)+S_{\alpha}(B|AC)+S_{\alpha}(C|AB),
	\end{equation}
	where e.g. $S_{\alpha}(A|BC)$ is the bipartite $\alpha$-R\'enyi entropy between tensor $A$ and the contraction of $B$ and $C$. We minimize $\mathrm{S3}_{\alpha}(A|B|C)$ by iterating over its terms and performing a single step of bipartite disentangling each time. It should be noted that each bipartite disentangler is here a three-body operator. After minimizing $\mathrm{S3}_{\alpha}(A|B|C)$ we get a triangle $ABC$ with minimal entanglement, and hence we will end up with a split-off slice that has minimal internal entanglement.
	
	In Appendix \ref{app_disent} we compare this tripartite disentangler with a direct 3D extension of the bipartite disentangler from Fig. \ref{fig:7}, where we replace each side of the yellow triangle in Fig. \ref{fig:9} by a pair of bipartite disentanglers.

	To continue the slice-splitting we absorb $A$ upwards and $B$ backwards, which is illustrated in Fig. \ref{fig:9}c and Fig. \ref{fig:8}b. We then temporarily combine the legs as indicated in Fig. \ref{fig:9}c and repeat the tetrahedron-splitting, yielding Fig. \ref{fig:8}c. Note that the new vertical red bond is enlarged to $D\chi$. With the orthogonality center at the top we perform a triangle-splitting to get Fig. \ref{fig:8}d, where we also truncated the enlarged vertical bonds back to $\chi$. This produces the first column of the new slice. By combining the tilted legs as in Fig. \ref{fig:9} we can repeat the splitting to move the TTN strand to the final column. We complete the slice-splitting by performing the column-splitting from Fig. \ref{fig:6}e-\ref{fig:6}h, resulting in Fig. \ref{fig:8}e. The transient slice is then absorbed into the next slice, after which the TTN strands have been successfully transferred, as illustrated in Fig. \ref{fig:8}f. Before continuing with the evolution we truncate the enlarged bonds.

	\subsubsection{Evolving the full network}\label{fullevol}
	
\begin{figure*}[!t]
		\begin{center}
			\includegraphics[width=\textwidth]{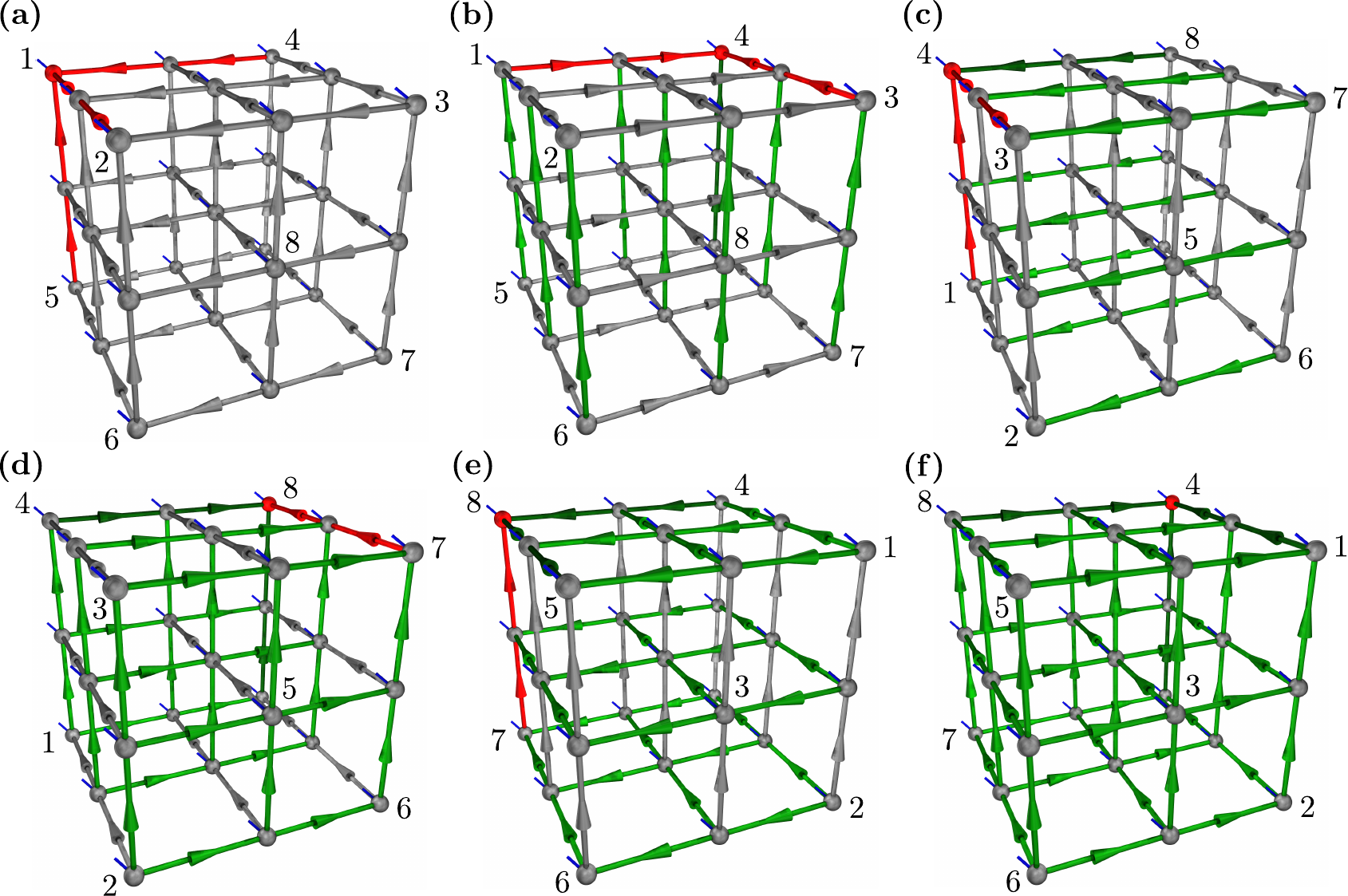}
		\end{center}
		\caption{A single iteration of $\mathrm{TEBD}^3$. Starting from the initial configuration in panel (a) we evolve all columns to get panel (b). Here the evolved gray bonds are colored green and the evolved red bonds are colored dark-green. To evolve the next set of columns we rotate the network as indicated by the numbering of the corners, yielding panel (c), which becomes panel (d) after evolving the new columns. To evolve the final set of columns we rotate the network to panel (e), which turns into panel (f) upon evolving the columns. With all bonds evolved this concludes a single iteration of $\mathrm{TEBD}^3$.}\label{evol_step}
	\end{figure*}

	With the machinery developed in the previous sections we can now perform a trotterized time-evolution where all truncation occurs at the orthogonality center, so that the local evolution is globally optimal. To illustrate this we consider the details of a single $\text{TEBD}^3$ iteration, starting from the network in Fig. \ref{evol_step}a. 
	
	We evolve the columns of the first slice, following the procedure in Fig. \ref{fig:6}, and subsequently move the TTN strands to the neighboring slice using the slice-splitting of Fig. \ref{fig:8}. Repeating these operations on the middle slice and afterwards evolving the final slice we end up with Fig. \ref{evol_step}b, where all columns have now been evolved as indicated by the green bonds.
	
	To evolve the rest of the bonds we rotate the network in Fig. \ref{evol_step}b such that the TTN again has the position from Fig. \ref{evol_step}a while the columns consist of non-evolved bonds. We have numbered the corners in Fig. \ref{evol_step} to show a possible way of doing this. After evolving the columns, yielding Fig. \ref{evol_step}d, we rotate the network to get \ref{evol_step}e, which becomes \ref{evol_step}f upon evolving the final columns. Since all bonds have now been evolved we have completed the iteration of $\text{TEBD}^3$.
	
    The $\text{TEBD}^3$ algorithm has two main sources of error: the error due to the multitude of truncated SVDs and the error due to the trotterization. In \cite{zaletel2019isometric} it was found that for $\text{TEBD}^2$ the column-splitting truncation error and trotterization error conspire to yield an energy-minimum in $\mathrm{d}\tau$-space. In Appendix \ref{app_trotter} we show that the same occurs for $\text{TEBD}^3$, now resulting from the interplay between the column- and slice-splitting truncation errors and the trotterization error.
	
	\subsubsection{Computational complexity}\label{compcost}
	
	The computational complexity of the $\text{TEBD}^3$ algorithm can be attributed to various operations, in particular to the SVDs that occur during these operations. Here we include only terms that are potentially leading in either $d$, $D$ or $\chi$. We moreover assumed that we perform full SVDs instead of partial SVDs in obtaining all estimates.
	
	Starting with the slice-splitting and subsequent absorption, we find that the tetrahedron-splitting has cost $\mathcal{O}(d D^{10} \chi^2 \min(d, D \chi^2))$ when it is performed on a tensor with the maximum amount of combined legs (see Fig. \ref{fig:9}). The truncation of the enlarged bonds after absorbing the split-off slice has cost $\mathcal{O}(d D \chi^6)$.
	
	Other contributions arise from the evolution of the bonds in a column, where the pairs of QR decompositions that precede gate-application have cost $\mathcal{O}(d \chi^5\min(d, \chi^2))$. These decompositions yield pairs of intermediate bonds with sizes $\eta$ and $\zeta$, which depend on the external bonds involved in the decompositions, so that the subsequent SVD has cost $\mathcal{O}(d^3\min(\eta\zeta^2,\eta^2\zeta))$. The pair that potentially yields the leading order in $D$ occurs in the bulk and has $\eta=\zeta=\chi \min(d,D^4)$.
	
	When simultaneously $d\leq\chi^2$ and $d\leq D^4$ the total cost reduces to $\mathcal{O}(d^2 D^{10} \chi^2) + \mathcal{O}(d D \chi^6) + \mathcal{O}(d^6 \chi^3)$. Because the costliest operations are performed roughly $N$ times we have the linear scaling $\mathcal{O}(N)$ in system size.

	\section{Benchmarking}
	
	\subsection{The benchmarking system}\label{benchsys}

	As a proof of principle we probe the accuracy of $\text{TEBD}^3$ for imaginary time evolution to find an isoTNS approximation for the many-body groundstate of a simple 3D quantum many-body system: the ferromagnetic transverse-field Ising model (TFIM) on a cubic lattice with OBC
	\begin{equation}
	H=-\sum_{\langle i j\rangle}\sigma^z_i\sigma^z_j -h\sum_i \sigma^x_i.\label{eq:H}
	\end{equation}
	Here $\vec{\sigma}_i\equiv(\sigma^x_i,\sigma^y_i,\sigma^z_i)$ corresponds to a spin-$1/2$ on site $i$, with $\sigma^{x,y,z}$ the usual Pauli matrices, and the two-body sum runs over nearest-neighbor pairs $\langle i, j\rangle$.
	
	Working in the $\sigma^z$-basis, we see from eq. (\ref{eq:H}) that the TFIM Hamiltonian becomes classical for $h\to0$ where it has a twofold degenerate ferromagnetic ground-state $\ket{\up\up\dots\up}$ and $\ket{\dw\dw\dots\dw}$.
	Quantum fluctuations are generated by the uniform magnetic transverse (i.e. along the $x$ direction) field of strength $h$. In the limit of strong fields $h\to\infty$, the ground state is unique and aligned with the field and a simple product state $\ket{\rightarrow\rightarrow\dots\rightarrow}$ in the $\sigma_x$ basis. Between these two limits, the competition between the $x$ and $z$ basis leads to a much more complex groundstate \cite{luitz_universal_2014} and to a quantum phase transition 
	as a function of $h$ between the ferromagnetic phase at $h<h_c$ and the polarized phase at $h>h_c$. The value of the critical field on the 3D simple cubic lattice was numerically estimated to be $h_c\approx5.15813(6)$ \cite{PhysRevE.66.066110}.
	
	Our main reason for choosing this model as a benchmark is its simplicity and the fact that it can be solved exactly using quantum Monte Carlo using the stochastic series expansion \cite{sandvik_quantum_1991,sandvik_generalization_1992}. Our two benchmark observables are the energy density $E/N=\langle H\rangle/N$ and the $x$-magnetization $m_x=\sum_i\langle\sigma_i^x\rangle/N$, which we compare to exact values in order to assess the accuracy of $\text{TEBD}^3$. 
	
	Quantum phase transitions are associated with a divergent correlation length and are therefore extremely challenging to study using tensor network methods. While a phase transition occurs at a singular point in the thermodynamic limit, a critical region is expected for finite systems.  In the case of OBC the region gets shifted towards $h<h_c$, which is especially pronounced for the smaller $L$, where the ratio of one- to two-body couplings is significantly larger. We expect the 3D isoTNS ansatz to be particularly challenged in this critical region, whereas its performance is likely to improve when progressing deeper into both phases, since the ground-state in both phases is a dressed product state.

	\subsection{The benchmarking procedure}
	
	We perform imaginary-time $\text{TEBD}^3$ propagation of the wave-function represented by an isoTNS with bond dimensions $(D,\chi)$ for the TFIM (\ref{eq:H}) with $L=3,4,10$ at various fields $h=0.5,1.0,\dots 8$ (across the critical point). For $L=3$ we use bond dimensions $D=2,4,6$ with $\chi=2D,4D,6D$, which allows us to observe the convergence of both $E/N$ and $m_x$ towards the exact values. For $L>3$ we are more constrained in our choice of $(D,\chi)$ due to the large cost as derived in Sec. \ref{compcost}. Hence for $L=4$ we use $D=2,3$ with $\chi=2D,4D,6D$, and to illustrate the capacity of $\text{TEBD}^3$ to reach large system sizes we consider a $(2,4)$ isoTNS for $L=10$.

	For the initial state we take a $\sigma_z$ product state $\ket{\psi_\text{ini}}=\ket{\up\up\dots\up}$, which we evolve in imaginary time until the energy density is converged. This means that we perform the operation
	\begin{equation}
	\ket{\psi_0} \approx \E^{-\beta \hat H } \ket{\up\up\dots\up},
	\label{eq:imag_te}
	\end{equation}
	which for $\text{TEBD}$ algorithms is done by trotterizing it into small timesteps $\D\tau$ as explained in Sec. \ref{sec:evolving3D}. We typically evolve to $\beta\gtrsim40$.
    
	As mentioned Sec. \ref{fullevol}, $\text{TEBD}^3$ has an energy minimum in $\D\tau$-space, so for each simulation we choose $\D\tau$ such that it coincides with the minimum. In Fig. \ref{fig:dt_dependency} of Appendix \ref{app_trotter} we show the $\D\tau$-minima for various $(D,\chi)$ on a $L=3$ lattice in the critical region. Similarly, we consider various R\'enyi entropy orders $\alpha$ for the disentangling and pick the one with lowest energy. It is usually between $\alpha=1/2$ and $\alpha=1$, although it varies across the phase diagram.
	
	The exact values for the energy density were obtained with the stochastic series expansion, for which we used the ALPS library \cite{Bauer_2011, ALBUQUERQUE20071187}. Here convergence to the ground-state was checked by a $\beta$-doubling scheme. For $L=3$ we also performed Lanczos exact diagonalization, providing exact energy density and additionally the $x$-magnetization of the ground-state at all $h$. 
	
	In order to provide a reference for the quality of our $\text{TEBD}^3$ results, we furthermore performed $\text{TEBD}^2$ calculations for the 2D TFIM using the algorithm detailed in \cite{zaletel2019isometric}. Here we chose a comparable system size of $5\times5$, for which exact Lanczos values are easily obtained.

	\subsection{Results}\label{results}
	
	In this section we present the results of our $\text{TEBD}^3$ benchmarks for the 3D TFIM ground-state.
	
\begin{figure}[]
		\begin{center}
			\includegraphics[width=\columnwidth]{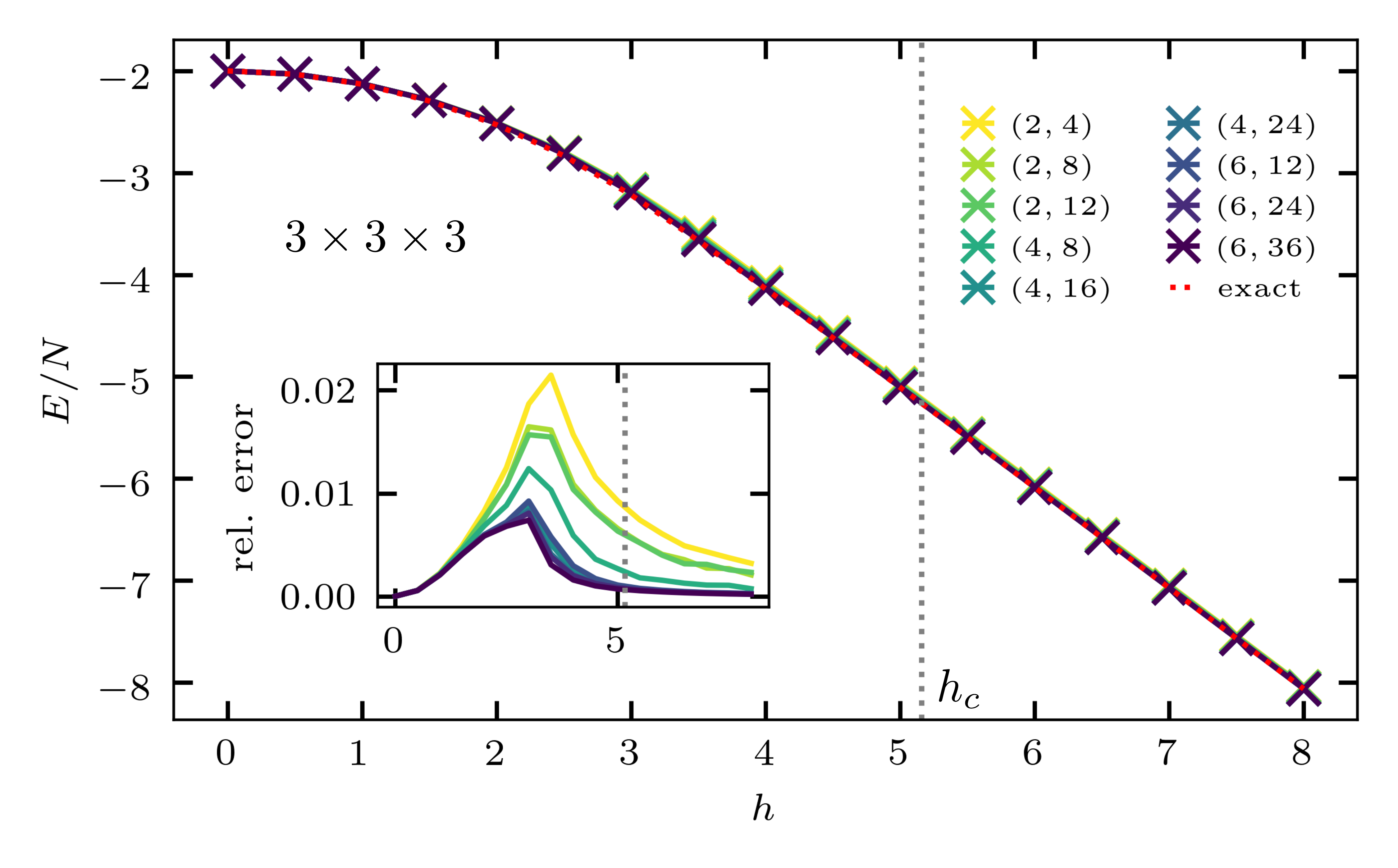}
			\includegraphics[width=\columnwidth]{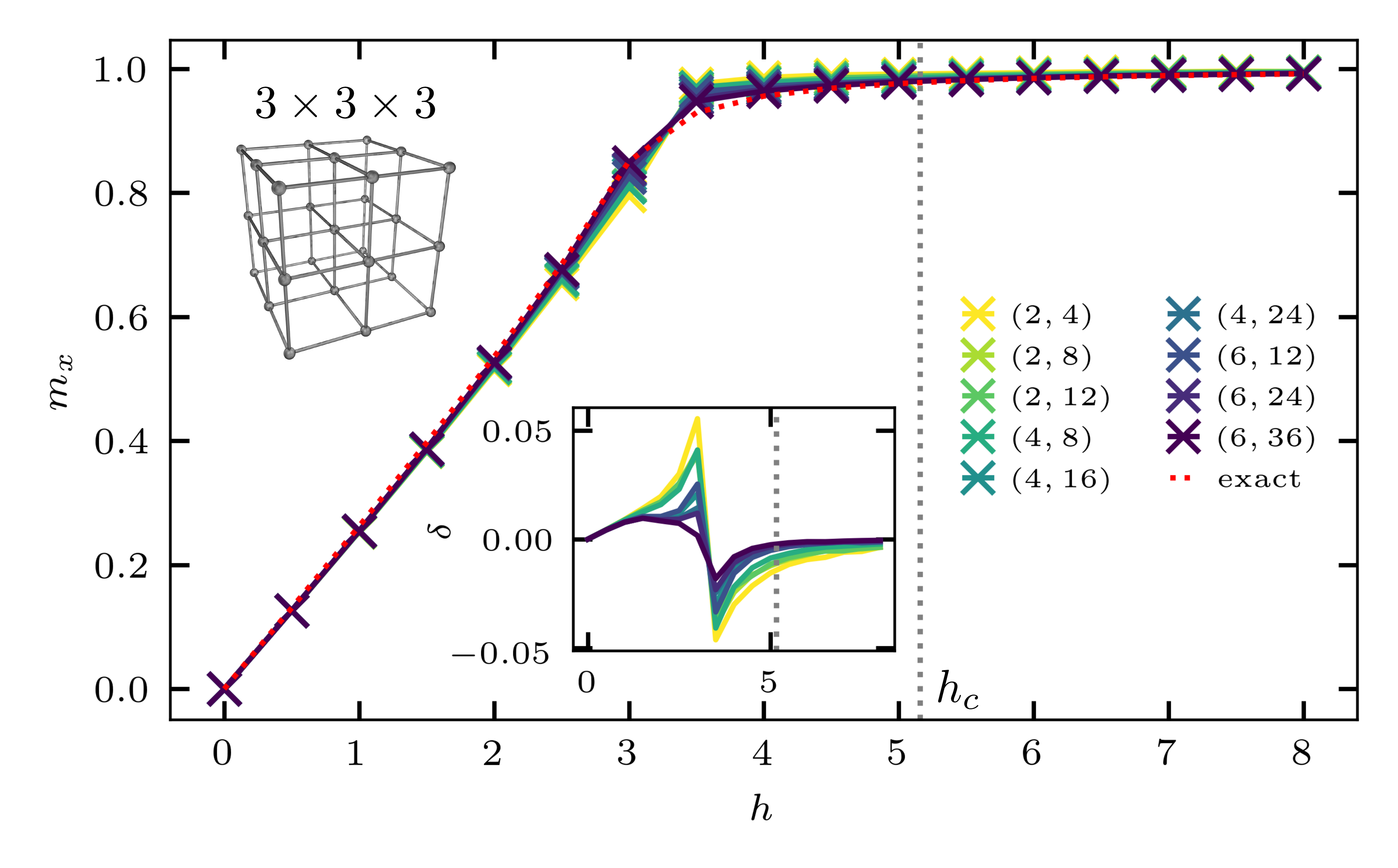}
			\caption{\textbf{Top:} The energy density $E/N$ and its accuracy relative to the exact Lanczos value for the $3\times3\times3$ cubic TFIM with OBC at various field strengths $h$ across the critical point. Each curve belongs to a different set of bond dimensions $(D,\chi)$, and in the main plot we show the exact values with a red dotted line. \textbf{Bottom:} The x-magnetization $m_x$ and its error $\delta=m_x^e-m_x$ relative to the exact Lanczos value $m_x^e$ for various field strengths around the critical point, for the same TFIM system as in the top panel.}\label{L3_accuracy}
		\end{center}
	\end{figure}

	We start off with $L=3$, for which the performance can be comprehensively probed. In Fig. \ref{L3_accuracy} we have plotted $E/N$ and $m_x$ across a range of $h$. Here the thermodynamic critical point is denoted by a gray dotted line, but as mentioned in Sec. \ref{benchsys} this point is spread into an extended region for finite systems, and furthermore shifted to smaller $h<h_c$ due to the use of OBC.

	The top panel of Fig. \ref{L3_accuracy} shows $E/N$ across a range of $h$ for various $(D,\chi)$, along with the exact Lanczos values as a red dotted line. The inset shows the relative error in $E/N$. Here we see that deep in the polarized and ferromagnetic phases the accuracy is high, whereas in the critical region around $h\approx3$ the performance is clearly worse. 

	For $D=2$ and $D=4$ we see that the accuracy is converged in $\chi$, meaning that $D$ must be increased (i.e. the column- and slice-splitting need higher fidelity) to further improve the performance. For $D=6$ it is not sure whether convergence is reached, meaning that convergence becomes slower as $D$ grows. On a related note, it can be seen that $(4,16)$ performs better than $(6,12)$, illustrating why large $D$ is only useful when combined with a few multiples larger $\chi$.
	
	The bottom panel of Fig. \ref{L3_accuracy} shows $m_x$ for various $h$, again together with the exact Lanczos values as a red dashed line. The inset contains the absolute error $\delta=m_x^e-m_x$ relative to the exact value $m_x^e$, which is chosen over the relative error since $m_x^e$ vanishes as $h\rightarrow0$. Deep in the polarized and ferromagnetic phases we see excellent agreement already for small bond dimensions, with the critical region clearly requiring larger tensors to get near the exact line.
	
\begin{figure}[]
		\begin{center}
   			\includegraphics[width=\columnwidth]{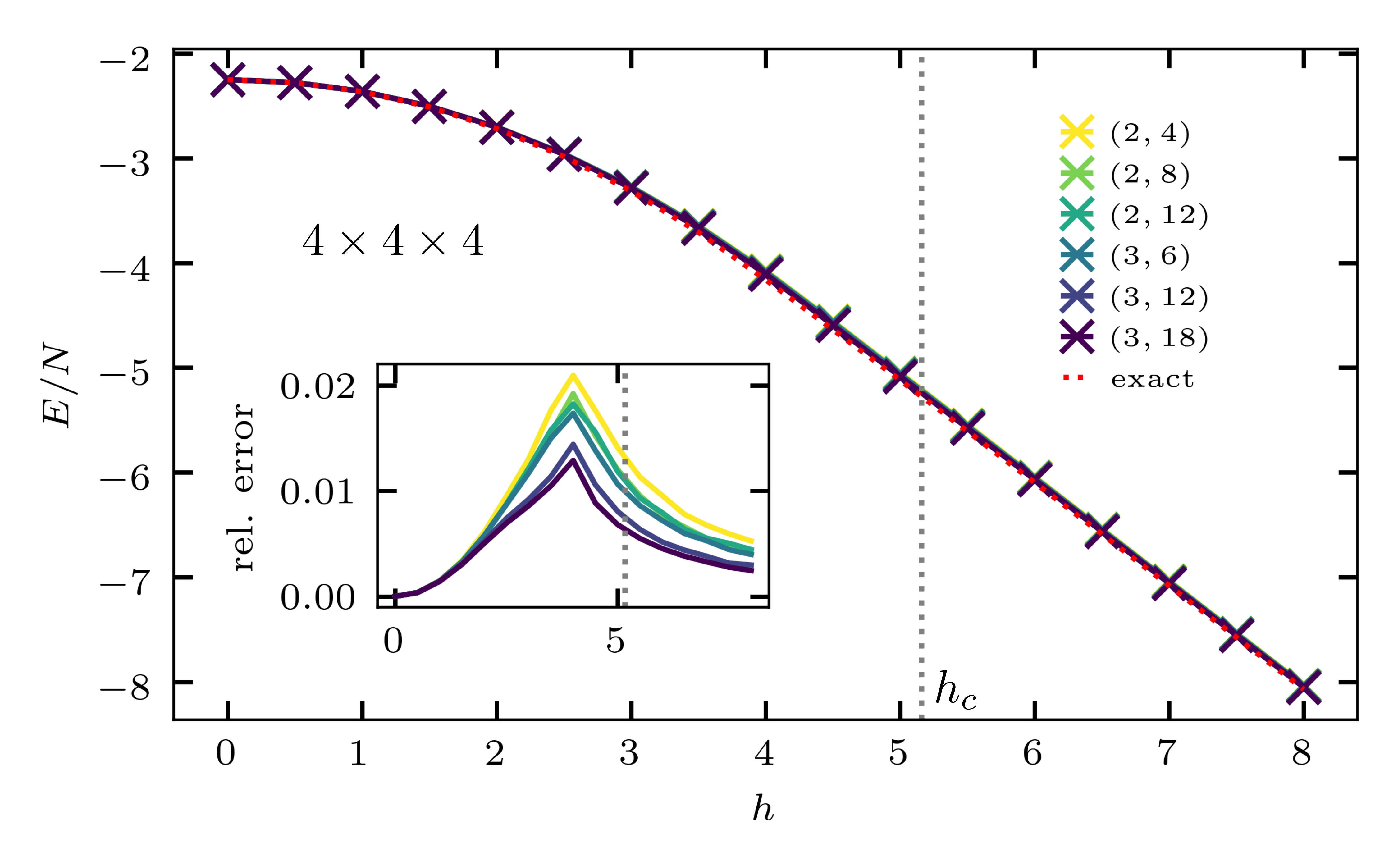}
			\includegraphics[width=\columnwidth]{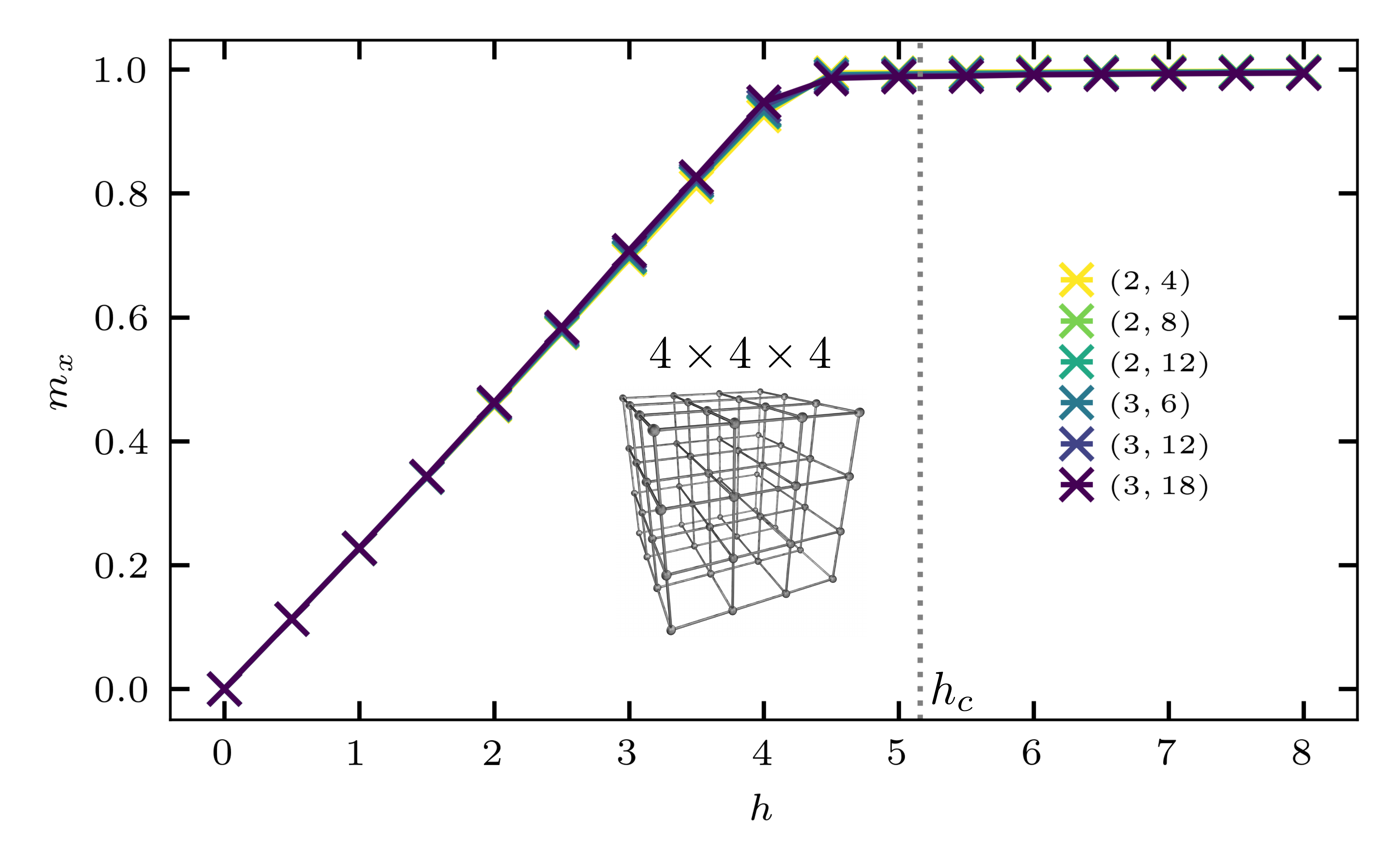}
			\caption{\textbf{Top:} The energy density $E/N$ and its accuracy relative to the exact QMC value for the $4\times4\times4$ cubic TFIM with OBC at various field strengths $h$ across the critical point. Each curve belongs to a different set of bond dimensions $(D,\chi)$, and in the main plot we show the exact values with a red dotted line. \textbf{Bottom:} The x-magnetization $m_x$ for various field strengths around the critical point, for the same TFIM system as in the top panel.}\label{L4_accuracy}
		\end{center}
	\end{figure}

    Next we consider $L=4$, for which we show the results in Fig. \ref{L4_accuracy}. The curves display similar behavior as for $L=3$, but due to the high computational cost we did not probe its performance beyond $D=3$. A noticeable difference is the sharper peak in energy accuracy, which is likely due to the shrinking critical region (that has moreover shifted to larger $h$). For $L=4$ the exact values were obtained via Monte Carlo, with the accompanying errors falling inside the line-width.

\begin{figure}[]
    	\begin{center}
    		\includegraphics[width=\columnwidth]{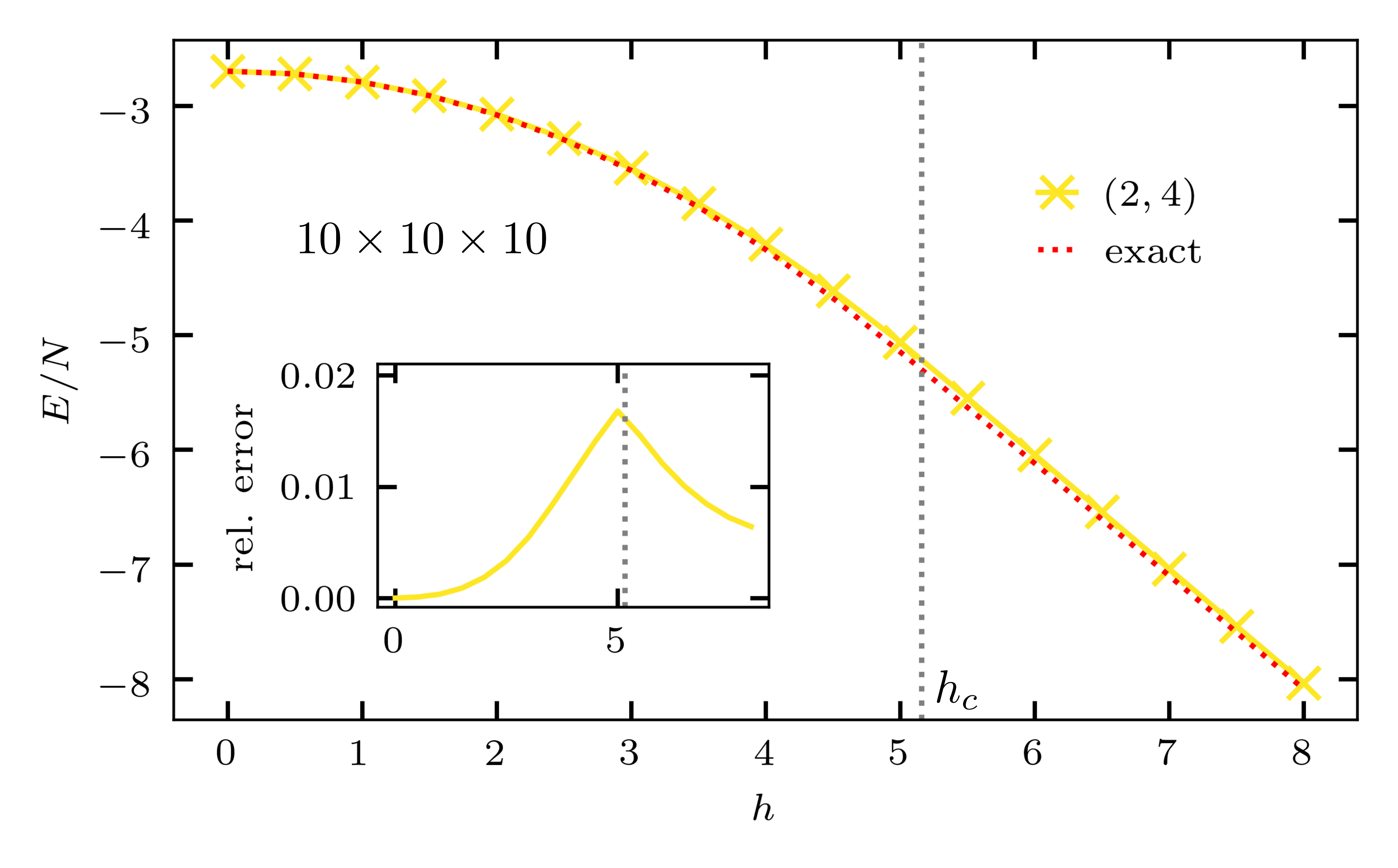}
    		\includegraphics[width=\columnwidth]{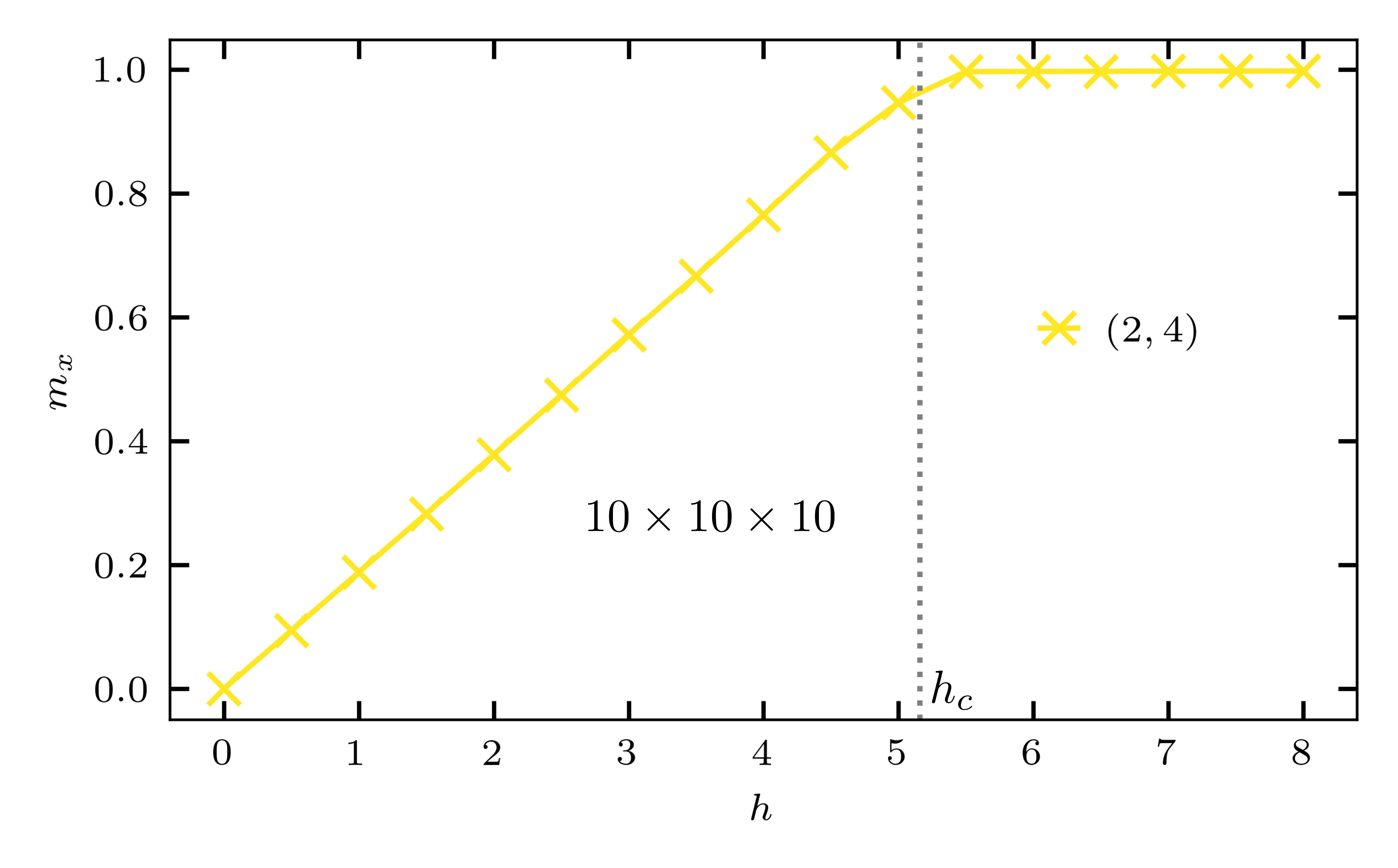}
    		\caption{\textbf{Top:} The energy density $E/N$ of the $(2,4)$ isoTNS and its accuracy relative to the exact QMC value for the $10\times10\times10$ cubic TFIM with OBC at various field strengths $h$ across the critical point. The exact values are shown as a red dotted line. \textbf{Bottom:} The x-magnetization $m_x$ for various field strengths around the critical point, for the same TFIM system as in the top panel.}\label{L10_accuracy}
    	\end{center}
    \end{figure}

	Now we consider $L=10$ (i.e. $N=10^3$) in Fig. \ref{L10_accuracy}. Here we were not able to go beyond $D=2$, which embodies the large difference in computational complexity between the bond dimensions and system size, as found in Sec. \ref{compcost}. We again observe a peak in relative error of $E/N$ at the critical region, which is now centered on the thermodynamic $h_c$. The $x$-magnetization is now also seen to saturate around $h_c$.
	
	\begin{figure}[]
		\begin{center}
			\includegraphics[width=\columnwidth]{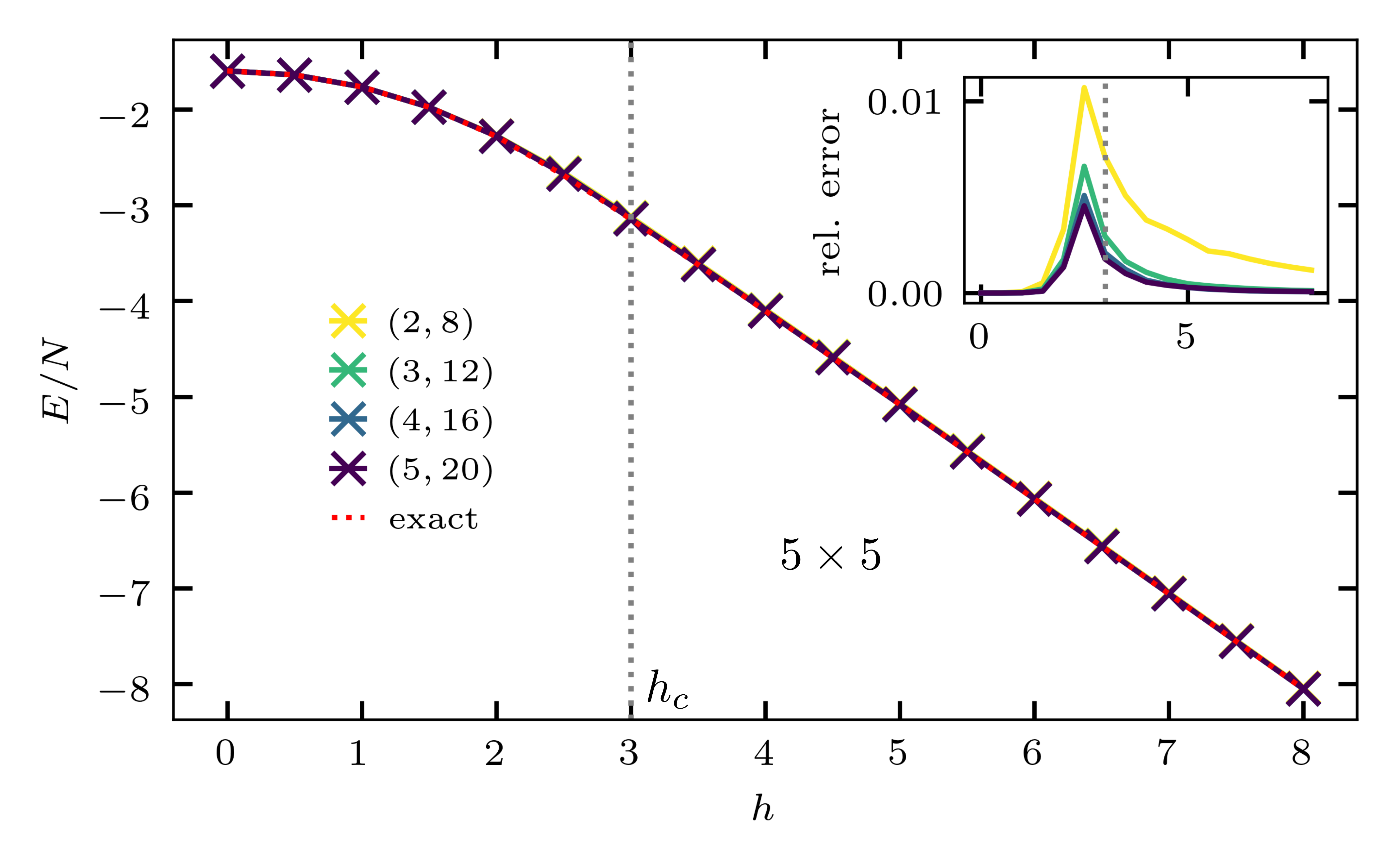}
			\includegraphics[width=0.94\columnwidth]{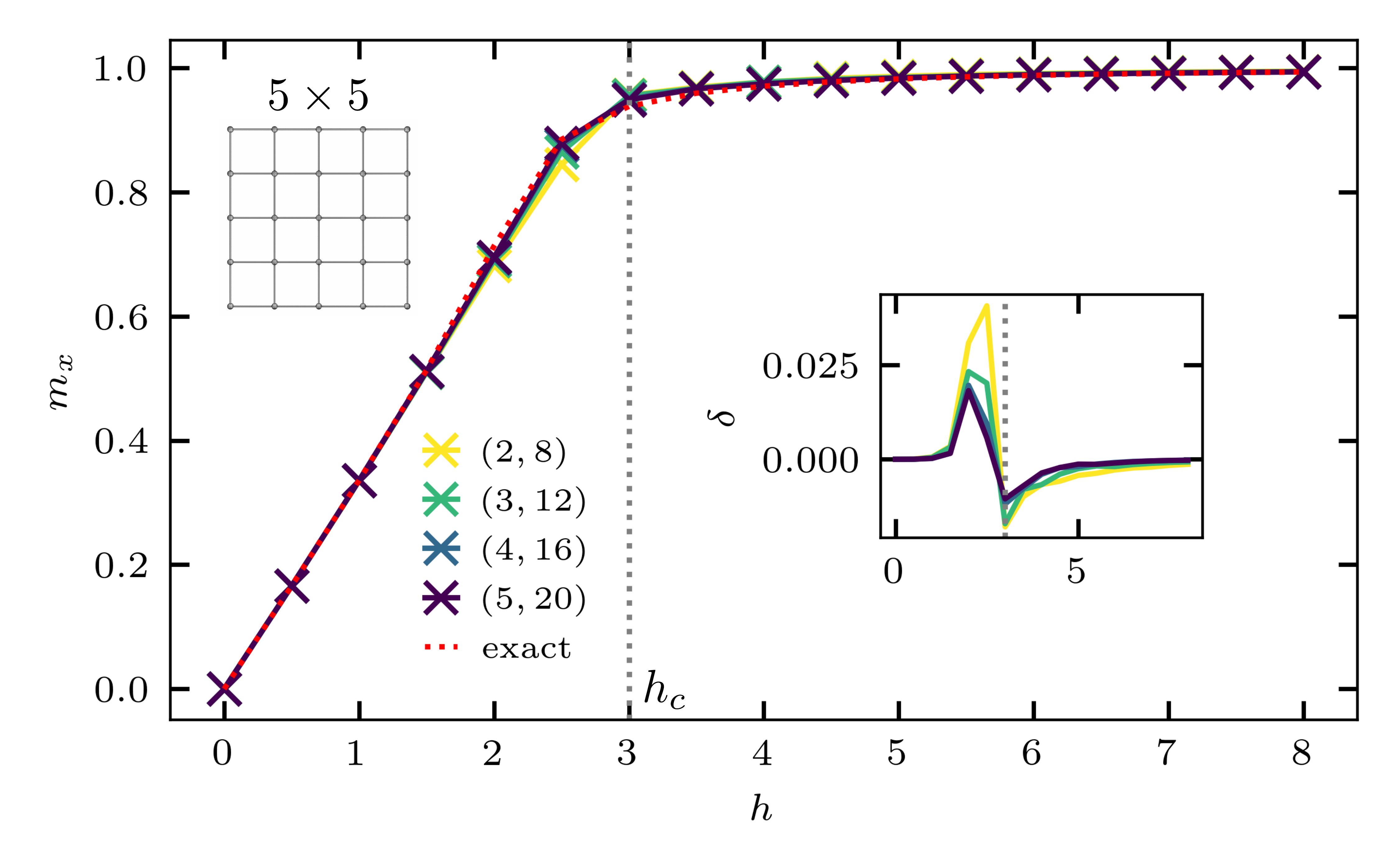}
			\caption{\textbf{Top:} The energy density $E/N$ and its accuracy relative to the exact Lanczos value for the $5\times5$ square TFIM with OBC at various field strengths $h$ across the critical point. Each curve belongs to a different set of bond dimensions $(D,\chi)$, and in the main plot we show the exact values with a red dotted line. \textbf{Bottom:} The x-magnetization $m_x$ and its error $\delta=m_x^e-m_x$ relative to the exact Lanczos value $m_x^e$ for various field strengths around the critical point, for the same TFIM system as in the top panel.}\label{2D_L5_accuracy}
		\end{center}
	\end{figure}
	
	Overall, the comparison of our results across different system sizes reveals an excellent representation of the ground-state wavefunction deep in the ferromagnetic phase, even for the relatively small $(D,\chi)$ which are reachable at large system sizes on current computers. The accuracy is slightly worse in the polarized phase, especially for the smaller $(D,\chi)$, but nonetheless high accuracies can be reached on relatively small lattices. As expected, the accuracy is worst in the critical region, but also here the performance is significantly improved upon increasing the bond dimensions, showing a clear trend toward the exact values both for the energy density and $x$-magnetization. 
	
	In order to put the $\text{TEBD}^3$ benchmarks into perspective we have also performed a $\text{TEBD}^2$ benchmark for the 2D TFIM on a $5\times5$ square lattice. It is clear that the simulation of a 2D system with $\text{TEBD}^2$ is easier than that of a 3D system with $\text{TEBD}^3$, since the 3D version involves more truncation. We therefore expect $\text{TEBD}^2$ to have higher accuracy for similar bond dimensions.
	
	In Fig. \ref{2D_L5_accuracy} we show $E/N$ and $m_x$ for multiple $(D,\chi)$ at various $h$ for the $5\times5$ TFIM. In the top panel we see that the relative error of $E/N$ is again smallest when deep in the ferromagnetic and polarized phases, with the critical region around $h_c\approx3$ again posing the biggest challenge. As expected, we see that $(4,16)$ performs better in 2D than in 3D, reaching just below $0.5\%$ compared to just below $1\%$ in Fig. \ref{L3_accuracy}. The same is apparent from the bottom panel, where we see that $(4,16)$ already closely matches the exact $m_x$ in the critical region, which is reached only with $(6,36)$ in Fig. \ref{L3_accuracy}.

	\section{Conclusion}
	
	We have introduced a method for the simulation of 3D quantum lattice models using a representation of the wavefunction as a 3D isometric tensor network state (isoTNS). Generalizing the method for 2D presented in \cite{zaletel2019isometric} we introduced a tetrahedral splitting and accompanying tripartite disentangling, such that optimal time-evolving block-decimation can be carried out in cubic 3D networks. We call the resulting evolution algorithm $\text{TEBD}^3$.
	
	Our systematic benchmark for the 3D transverse field Ising model in the full range of transverse fields across the critical point reveals that our method yields accurate results, and that the systematic error incurring from finite bond dimensions can be controlled systematically by increasing $(D,\chi)$. This behavior is identical to what is observed in $\text{TEBD}^2$. The regime close to the critical point is particularly challenging and requires larger bond dimensions, beyond the capacity of our computers for large systems.
	
	While imaginary time-evolution using $\text{TEBD}^3$ is arguably the simplest method for finding the groundstate of a quantum many-body system, it is known even in 1D that it is not optimal and that local variational energy minimization (e.g. DMRG in 1D) is far more efficient. We expect a similar behavior for isoTNS in higher dimensions and it is possible that our results can be further improved by the formulation of a DMRG analog for 3D isoTNS, a direction which we leave for future study.

	\section{Acknowledgments}
	
	We are grateful to M. Zaletel and F. Pollmann for useful suggestions to further improve isoTNS. We acknowledge financial support from the Deutsche Forschungsgemeinschaft through SFB 1143 (project-id 247310070).

	\section{Appendix}

	\subsection{Bipartite versus tripartite disentangling} \label{app_disent}
	
\begin{figure}[h]
		\begin{center}
			\includegraphics[width=\columnwidth]{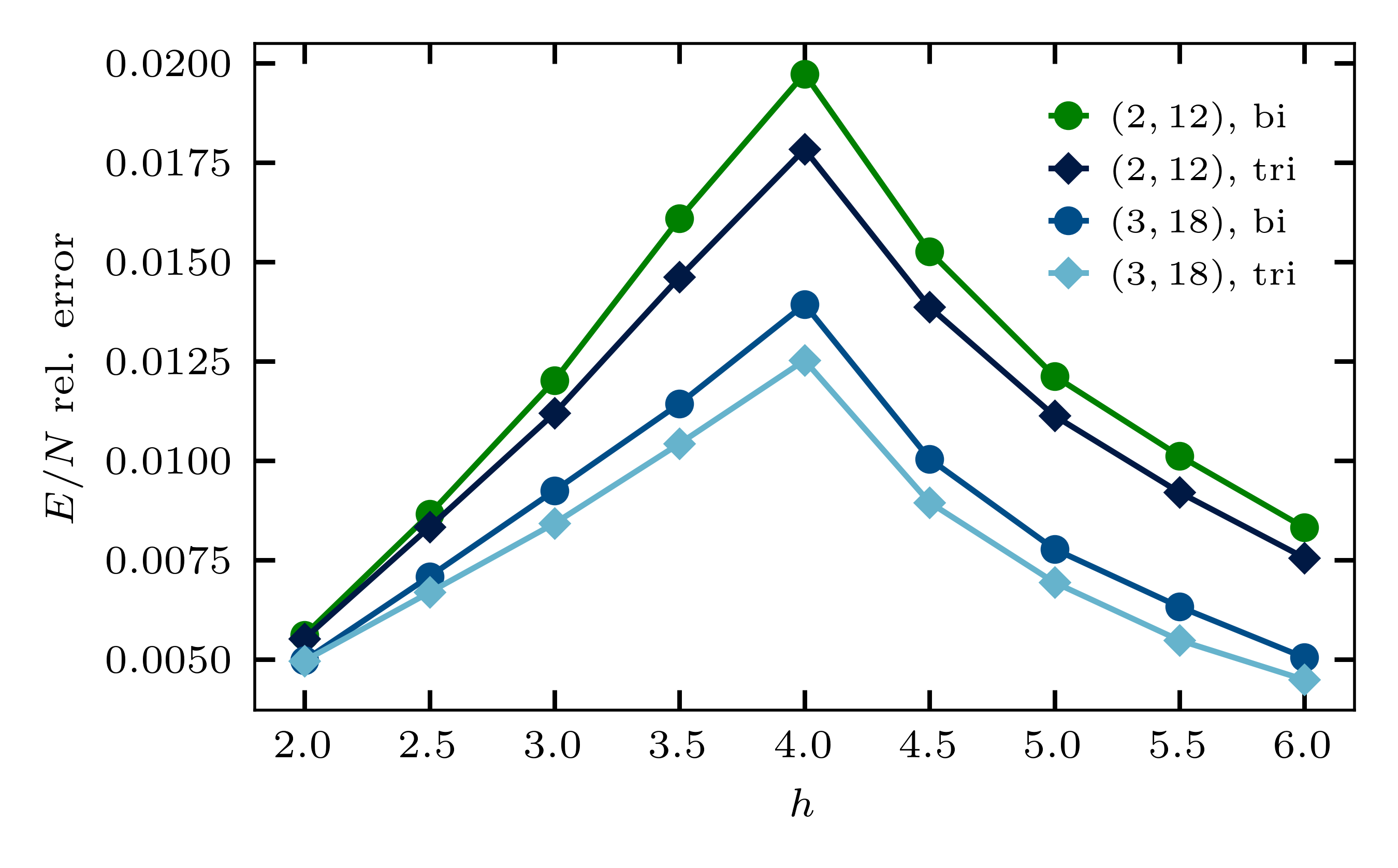}
		\end{center}
		\caption{A comparison of the bipartite and tripartite disentanglers, showing the energy accuracy for the $4\times4\times4$ TFIM ground-state at various $h$ and two $(D,\chi)$. We see that the improvement in using a tripartite over a bipartite disentangler is most significant in the critical region.}\label{fig:disents}
	\end{figure}

	To quantify the difference in performance between the bipartite and tripartite 3D disentanglers from Section \ref{split_slice} we use each to calculate the ground-state energy density of the $4\times4\times4$ TFIM at various $h$. In Fig. \ref{fig:disents} we show the performance of the isoTNS configurations $(2,12)$ and $(3,15)$. Here it is clear that the improvement in accuracy when using tripartite over bipartite disentanglers is largest in the critical region $h\in[3,5]$, whereas the improvement becomes progressively smaller outside of this region. 
	
	Because the chosen $(D,\chi)$ are converged in $\chi$ we cannot improve the bipartite curves by further increasing the TTN's bond dimension. This illustrates that the quality of the tetrahedron-splitting, and hence of the slice-splitting, cannot be compensated by only improving the TTN. This is easily understood if we recognize that the slice-splitting is a major component of $\text{TEBD}^3$, since it serves to transfer the TTN to the next slice with minimal information loss, and that this part of the algorithm is mainly controlled by $D$ and not $\chi$ (see Fig. \ref{fig:9}). In particular, even though a larger $\chi$ might improve the time-evolution on a fixed TTN, this gain is lost when transferring the TTN.
	
	\subsection{Minima in $\text{d}\tau$ space} \label{app_trotter}
	
	As noted in \cite{zaletel2019isometric}, the error due to the multitude of truncated SVDs and the error due to the trotterization combine to yield an energy-minimum in $\mathrm{d}\tau$-space for $\text{TEBD}^2$. Here we show that this also occurs for $\text{TEBD}^3$ and we will furthermore illustrate the difference between first- and second-order trotterization.
	
\begin{figure}[h]
		\begin{center}
			\includegraphics[width=\columnwidth]{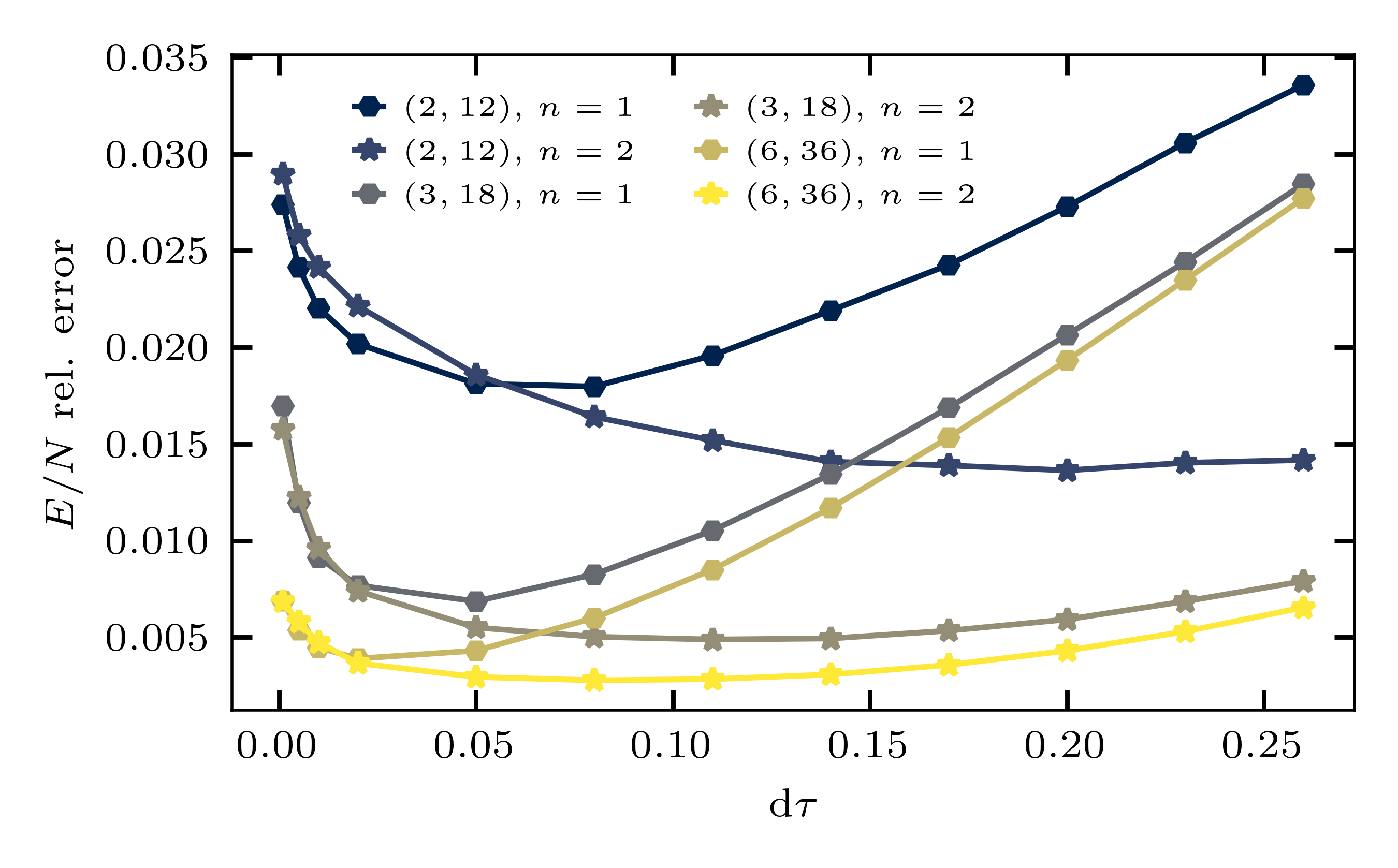}
		\end{center}
		\caption{The $\D\tau$-dependency of the relative error in energy density of the $3\times3\times3$ TFIM at $h=3.5$, for various isoTNS configurations $(D,\chi)$.}\label{fig:dt_dependency}
	\end{figure}

	In Fig. \ref{fig:dt_dependency} we show the relative error in energy density for the ground state of the $3\times3\times3$ TFIM at $h=3.5$, at multiple points in $\D\tau$-space. For each $(D,\chi)$ we plot both the first-order ($n=1$) and second-order ($n=2$) trotterization results, from which we see that for all considered cases the $n=2$ minimum lies below the $n=1$ minimum. We can also see that the minima shift to lower $\D\tau$ as we increase $D$, for both $n=1$ and $n=2$, which is in accordance with the findings in \cite{zaletel2019isometric} for $\text{TEBD}^2$.
	
	\bibliography{tebd3.bbl}

\end{document}